\documentclass{emulateapj}
\usepackage{natbib}
\usepackage{multirow}
\shorttitle{HALOGAS:  NGC 4565}
\shortauthors{Zschaechner et al.}

\begin{document}

\title{HALOGAS: {\sc H\,i} Observations and Modeling of the Nearby Edge-on Spiral Galaxy NGC 4565} 
\author{\sc Laura K. Zschaechner,\altaffilmark{1} Richard J. Rand,\altaffilmark{1}, George H. Heald\altaffilmark{2},\\ Gianfranco Gentile \altaffilmark{3,4} \& Gyula J\'{o}zsa \altaffilmark{2}}
\altaffiltext{1}{Department of Physics and Astronomy, University of New Mexico, 1919 Lomas Blvd NE, Albuquerque, New Mexico 87131-1156; zschaech@unm.edu, rjr@phys.unm.edu}
\altaffiltext{2}{Netherlands Institute for Radio Astronomy (ASTRON), Postbus 2, 7990 AA Dwingeloo, the Netherlands; heald@astron.nl, jozsa@astron.nl}
\altaffiltext{3}{Sterrenkundig Observatorium, Ghent University, Krijgslaan 281, S9, 9000 Ghent, Belgium; gianfranco.gentile@ugent.be}
\altaffiltext{4}{Astrophysical Institute, Vrije Universiteit Brussel, Pleinlaan 2, 1050 Brussels, Belgium}
\slugcomment{Accepted to Astrophysical Journal on October 1, 2012}
\begin{abstract}

\par
     We present 21-cm observations and models of the neutral hydrogen in NGC 4565, a nearby, edge-on spiral galaxy, as part of the Westerbork Hydrogen Accretion in LOcal GAlaxieS (HALOGAS) survey.  These models provide insight concerning both the morphology and kinematics of {\sc H\,i} above, as well as within, the disk.  NGC 4565 exhibits a distinctly warped and asymmetric disk with a flaring layer.  Our modeling provides no evidence for a massive, extended {\sc H\,i} halo.  We see evidence for a bar and associated radial motions.  Additionally, there are indications of radial motions within the disk, possibly associated with a ring of higher density.  We see a substantial decrease in rotational velocity with height above the plane of the disk (a lag) of $-$ 40 $^{+5}_{-20}$ km s$^{-1}$ kpc$^{-1}$  and $-$ 30 $^{+5}_{-30}$ km s$^{-1}$ kpc$^{-1}$ in the approaching and receding halves, respectively.  This lag is only seen within the inner {\raise.17ex\hbox{$\scriptstyle\sim$}}4.75' (14.9 kpc) on the approaching half and {\raise.17ex\hbox{$\scriptstyle\sim$}}4.25' (13.4 kpc) on the receding, making this a radially shallowing lag, which is now seen in the {\sc H\,i} layers of several galaxies.  When comparing results for NGC 4565 and those for other galaxies, there are tentative indications of high star formation rate per unit area being associated with the presence of a halo. Finally, {\sc H\,i} is found in two companion galaxies, one of which is clearly interacting with NGC 4565.

\end{abstract}


\section{Introduction} \label{Section1}

\par
     Understanding the origins and evolution of extra-planar gas is key to interpreting how spiral galaxies are affected by their environments.  Whether galaxies exist in dense clusters or groups, or are relatively isolated, extra-planar gas is the bridge between galaxy disks and the intergalactic medium (IGM).  Is this gas ejected from the plane of the disk, driven by star formation as described in the galactic fountain (\citealt{1976ApJ...205..762S}, \citealt{1980ApJ...236..577B}) and chimney \citep{1989ApJ...345..372N} models?  Does some of it result from accretion from external sources (e.g.\ \citealt{2009ApJ...700L...1K}) such as companion galaxies or the IGM itself?  If so, how would these internal and external components interact with each other?  To answer these questions, a thorough study of extra-planar gas morphology and kinematics in a large number of galaxies is necessary.

\par
     Extra-planar components including hot gas (e.g.\ \citealt{2006A&A...448...43T}), relativistic particles (e.g.\ \citealt{1999AJ....117.2102I}), dust (e.g.\ \citealt{1999AJ....117.2077H}), ionized hydrogen (e.g.\ \citealt{1996ApJ...462..712R}; \citealt{2003A&A...406..493R}) as well as neutral hydrogen ({\sc H\,i}) (e.g.\ \citealt{1997ApJ...491..140S}, \citealt{2007AJ....134.1019O}) have been observed.   For all except {\sc H\,i} there is a correlation between their presence and star formation in the disk, both in localized regions as well as globally.  This indicates disk-halo flows akin to those in the aforementioned galactic fountain and chimney models as likely origins.  However, such trends have not been reported for {\sc H\,i}, and in fact there is evidence that at least some extra-planar {\sc H\,i} has an external origin in the Milky Way \citep{1997ARA&A..35..217W} and nearby galaxies \citep{2008A&ARv..15..189S}.  This is an issue that motivates this work.

\par
     In addition to probing the origins of different halo components through their morphology, we may also gain information through understanding their kinematics.  In galactic fountain-type models, gas is launched from the plane of the disk and moves to larger radii.  In order to conserve angular momentum, there must be a decrease in rotational velocity with height.  This vertical gradient in rotational velocity is referred to as a lag.  However, other physical processes may affect the magnitude of such a lag (see below); hence, measurement of lags should help to constrain the physics of extra-planar components, and may shed light on important elements of galaxy evolution.

\par
     \citet{2007ApJ...663..933H} measured lags in three extra-planar Diffuse Ionized Gas (DIG) layers, resulting in a trend of low star formation rates (SFR) with steeper lags.  If {\sc H\,i} and DIG layers share similar kinematics and the \citet{2007ApJ...663..933H} trend holds, then, the same should be expected for {\sc H\,i}.

\par
  Simple dynamical models of disk-halo cycling, where only ballistic effects are considered (\citealt{2002ApJ...578...98C}, \citealt{2006MNRAS.366..449F}), have produced lags significantly shallower than those observed. These discrepancies between models and observations indicate that additional factors must be considered. Firstly, internal complications such as those produced by extra-planar pressure gradients or magnetic tension \citep{2002ASPC..276..201B} have yet to be examined thoroughly, but could potentially explain these discrepancies without resorting to external factors. 

\par
     Explanations involving infalling gas such as interaction of disk-halo cycled gas with a possible hot, low angular momentum, low metallicity halo predicted by cosmological simulations \citep{2011MNRAS.415.1534M} and the hydrodynamic simulation of disk formation in \citet{2006MNRAS.370.1612K} have successfully reproduced the observed lag in NGC 891.  

\par
    However, while much attention has been focused on results for NGC 891, observed lags differ among the three galaxies considered in \citet{2007ApJ...663..933H} and one would like to understand the physical cause of such variations.  More relevant here, extensive modeling of extra-planar {\sc H\,i} kinematics has until recently been carried out for only a small number of edge-on galaxies, using observations with a range of resolutions and sensitivities.  These include NGC 891 \citep{2007AJ....134.1019O}, where DIG and {\sc H\,i} lags agree, pointing to similar origins for both components, and other galaxies summarized in $\S$~\ref{Section6.4}.  Furthermore, the trend of steeper lags with lower SFRs found by \citet{2007ApJ...663..933H} needs to be further tested, while variation of lags \textit{within} extra-planar gas layers, such as radial gradients, may also constrain their origin.  With the aforementioned evidence for an external source for at least some extra-planar {\sc H\,i}, it is imperative to understand how kinematic information may constrain these two possible origins of extra-planar gas.  

\par
     The Westerbork Hydrogen Accretion in LOcal GAlaxies (HALOGAS) survey \citep{2011A&A...526A.118H},  which targets 22 edge-on or moderately inclined spiral galaxies for deep, uniform 10$\times$12 hour observations using the Westerbork Synthesis Radio Telescope (WSRT), will greatly increase this number. This survey will also allow us to establish whether there is any connection between extra-planar {\sc H\,i} and star formation, as well as the degree to which contributions from external origins are relevant.  A primary goal of HALOGAS is to estimate the rate of {\sc H\,i} accretion in spiral galaxies.

\par
    Aside from the characterization of extra-planar gas, one goal of HALOGAS is to investigate morphological features such as bars and rings as well as their kinematics. These features will be discussed throughout this paper.

\subsection{NGC 4565} \label{Section1.1}
\par
    NGC 4565 is classified as a SAb galaxy \citep{1991trcb.book.....D} with a SFR of 0.67 M$_{\odot}$ yr$^{-1}$ calculated in \citet{2012A&A...544C...1H} using total infrared (TIR) flux measurements.  This places it near the low star-forming end of the HALOGAS sample.  However, based on its rotation curve, it is among the most massive, which would presumably hinder gas being ejected to large heights above the disk.  This combination gives it substantial importance in determining overall trends involving extra-planar {\sc H\,i} and star formation.  

\par
     At a distance of 10.8 Mpc \citep{2011A&A...526A.118H}, NGC 4565 is relatively nearby, and among the best studied edge-on spirals.  It has been observed in optical continuum (e.g.\ \citealt{2010ApJ...715L.176K}), H$\alpha$ \citep{1992ApJ...396...97R}, radio continuum \citep{2008BASIP..25...67K}, CO line emission \citep{1996A&A...310..725N}, and X-rays \citep{1996A&A...305...74V}.

\par
    In the optical, NGC 4565 is seen to have a box-shaped bulge indicating a bar \citep{2010ApJ...715L.176K}.  Additionally, \citet{1996A&A...310..725N} found radial motions in CO emission along the minor axis, which they suggested could be due to a bar or a spiral density wave.  We see evidence for the bar in {\sc H\,i} as will be discussed in $\S$~\ref{Section3}. 

\par
    \citet{2008BASIP..25...67K} detect a radio continuum halo extending up to 3 kpc.  In contrast, \citet{1992ApJ...396...97R} found that NGC 4565 was lacking a smooth DIG halo, which is consistent with its low star formation rate.

\par
    Unexpectedly, substantial X-ray emission, generally not seen in galaxies without enhanced star formation activity, was detected both within and above the disk \citep{1996A&A...305...74V}.  The nature of this X-ray emission is still unknown.

\par
    Finally, {\sc H\,i} observations of NGC 4565 were previously performed using the Very Large Array (VLA), but detailed models were not created from those data \citep{1991AJ....102...48R}.  However, a comprehensive description was provided in that paper, mentioning several features we now see in greater detail, and with a higher level of confidence.  Additional WSRT data were taken \citep{2005A&A...432..475D}, which were further interpreted in \citet{2005ASPC..331..139V}.  These WSRT data are supplemented with the HALOGAS data to produce the final dataset presented here.  ~\ref{tbl-1} summarizes parameters for NGC 4565.

\begin{deluxetable*}{lrr}
\tabletypesize{\scriptsize}
\tablecaption{NGC 4565 Parameters\label{tbl-1}}
\tablewidth{0pt}
\tablehead
{
\colhead{Parameter} &
\colhead{Value}&
\colhead{Reference}\\
}
\startdata
\phd Distance (Mpc) &10.8\tablenotemark{a} &Heald et al. (2011)\\ 
\phd Systemic velocity (km s$^{-1}$)&1220 &This work\\
\phd Inclination &87.5$\,^{\circ}$ &This work\\
\phd SFR (M$_{\odot}$ yr$^{-1}$) &0.67 &Heald et al. (2011)\\ 
\phd Morphological Type &SAb\tablenotemark{b} & \citet{1991trcb.book.....D} \\
\phd Kinematic Center $\alpha$ (J2000.0) &  12h 36m 21s &This work\\
\phd Kinematic Center $\delta$ (J2000.0) & 25d 59m 10s &This work\\
\phd D$_{25}$ (kpc) &16.2& \citet{1991trcb.book.....D}\\
\phd Total Atomic Gas Mass ($10^{9}M_{\odot}$) & 9.9 \tablenotemark{c}&This work\\
\enddata
\tablenotetext{a}{Distance is the median value of distances found on the NED database, excluding those obtained using the Tully-Fisher relation.}
\tablenotetext{b}{The A classification indicating no bar is likely erroneous based on Neininger et al. 1996, Kormendy \& Barentine 2010, and this work.}
\tablenotetext{c}{Includes neutral He via a multiplying factor of 1.36.}
\end{deluxetable*}

\section{Observations and Data Reduction} \label{Section2}

\par
     Here we provide a brief explanation of the observations and data reduction process.  A detailed description, which applies to all of the HALOGAS galaxies, may be found in \citet{2011A&A...526A.118H}.      

\par
     NGC 4565 was observed using the Westerbork Synthesis Radio Telescope (WSRT) by \citet{2005A&A...432..475D} as well as through the HALOGAS survey.  The archival observations were done in the Traditional WSRT configuration, using four distinct spacings between the fixed and movable antennas, starting at 36 m and incremented by 18 m.  Observations obtained as part of HALOGAS were in the Maxi-short configuration with baselines ranging from 36 m to 2.7 km in order to maximize sensitivity to faint, extended emission.  Of the 10 fixed antennas, 9 were used and spaced at 144 m intervals on a regular grid. The total bandwidth is 10 MHz with 1024 channels and two linear polarizations. The total observing time is $10\times12$ hours, and observing dates are listed in Table~\ref{tbl-2}. 

\begin{deluxetable}{lr}
\tabletypesize{\scriptsize}
\tablecaption{Observational and Instrumental Parameters  \label{tbl-2}}
\tablewidth{0pt}
\tablehead
{
\colhead{Parameter} &
\colhead{Value}\\
}
\startdata
\phd Observation Dates $-$ HALOGAS&2009 Mar 10-11\\
\phd &2009 Mar 11-12\\
\phd &2009 Mar 15-16\\
\phd &2009 Apr 1-2\\
\phd &2009 Apr 3-4\\
\phd &2009 Apr 12-13\\
\phd &2009 Apr 24\\
\phd &2009 Apr 27-28\\
\phd Observation Dates $-$ Dahlem&2003 Feb 16-17\\
\phd &2003 Apr 14-15\\
\phd &2003 May 14-15\\
\phd &2003 May 19-20\\
\phd Pointing Center &12h 34m 21.071s\\
\phd &25d 59m 13.50s\\ 
\phd Number of channels &148\\
\phd Velocity Resolution &4.12 km s$^{-1}$\\
\phd Full Resolution Beam Size &24.27$\times$11.5"\\
\phd &1270$\times$600 pc\\
\phd RMS Noise $-$ 1 Channel (24.27$\times$11.5")&0.31 mJy bm$^{-1}$\\
\phd RMS Noise $-$ 1 Channel (44$\times$34")&0.24 mJy bm$^{-1}$\\
\enddata
\end{deluxetable} 

\par
  Data reduction was performed in Miriad \citep{1995ASPC...77..433S}. Images were created with a variety of weighting schemes, creating multiple data cubes.  Clark deconvolution \citep{1980A&A....89..377C} was performed using mask regions defined on the basis of unsmoothed versions of the image cube.  Offline Hanning smoothing led to a final velocity resolution of 4.12 km s$^{-1}$ in a single channel.  The 1$\sigma$ rms noise in a single channel of the full resolution cube is 0.31 mJy bm$^{-1}$, corresponding to a column density of  $N_{HI}=5.1\times10^{18}$ cm$^{-2}$.

\par
    To optimize our resolution, we use a cube with a robust parameter of $-$2. The beam for this cube is 24.27$\times$11.45"; 1" = 52.4 pc (1270$\times$600 pc) with a position angle of $-0.47\,^{\circ}$.  By using this cube, we sacrifice some sensitivity to faint extended emission.  However, we later examine and apply the same models to a second cube where a Gaussian {\it uv} taper yielding 44$\times$34" resolution is used.  Primary beam correction has been applied to both cubes using the Miriad task {\tt linmos}.  In the case of moment maps, which are created using methods described in \citet{2012MNRAS.422.1835S}, this is done in the last step.

\section{The Data} \label{Section3}

\par
\begin{figure}
\epsscale{1}
\figurenum{1}
\plotone{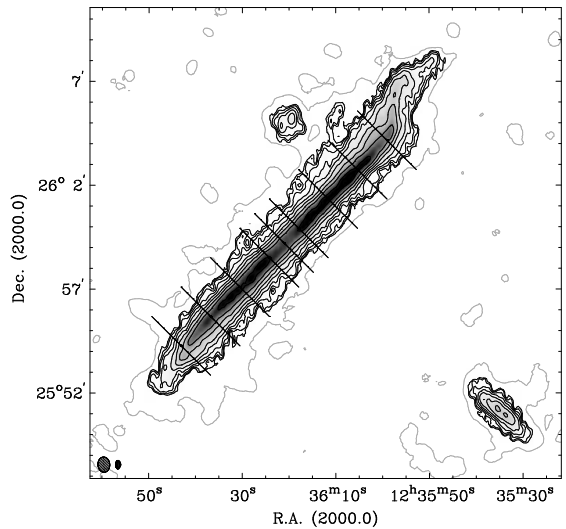}
\caption{The {\sc H\,i} zeroth-moment map displaying the full-resolution (black contours and grayscale) and smoothed $44"\times34"$ (gray contour) cubes.  Full resolution contours begin at $3.0\times10^{19}$ $\mathrm{cm^{-2}}$ and increase by factors of 2.  The smoothed contour is at a level of $1.1\times10^{19}$ $\mathrm{cm^{-2}}$.  Both beams are shown in the lower left-hand corner.  Two companion galaxies may be seen, NGC 4562 in the southwest corner and IC 3571 north of the center.  Note the strong, asymmetric warp across the line of sight as well as the connection between NGC 4565 and IC 3571.  Diagonal lines parallel to the minor axis indicate slice locations shown in Figure~\ref{fig3}.  \label{fig1}}
\end{figure}

    Upon initial inspection of the zeroth moment map (Figure~\ref{fig1}), {\sc H\,i} emission can consistently (i.e.\ along most of the disk) be seen up to heights of 1.5-1.75' in the full-resolution and smoothed cubes respectively.  As will be shown in $\S$~\ref{Section4}, this does not mean that an extended {\sc H\,i} halo is necessary to fit the data.  Also evident in the zeroth moment map is an asymmetric warp component across the line of sight.  This warp is most prevalent in the receding (NW) half of the galaxy, where the disk tilts upward and then flattens at large radii.  In the faint emission, a connection is seen between NGC 4565 and the companion in the northeast quadrant (IC 3571), which was also noted by \citet{2005ASPC..331..139V}.  This interaction may contribute to the asymmetry of the warp.

\par
\begin{figure}
\epsscale{1}
\figurenum{2}
\plotone{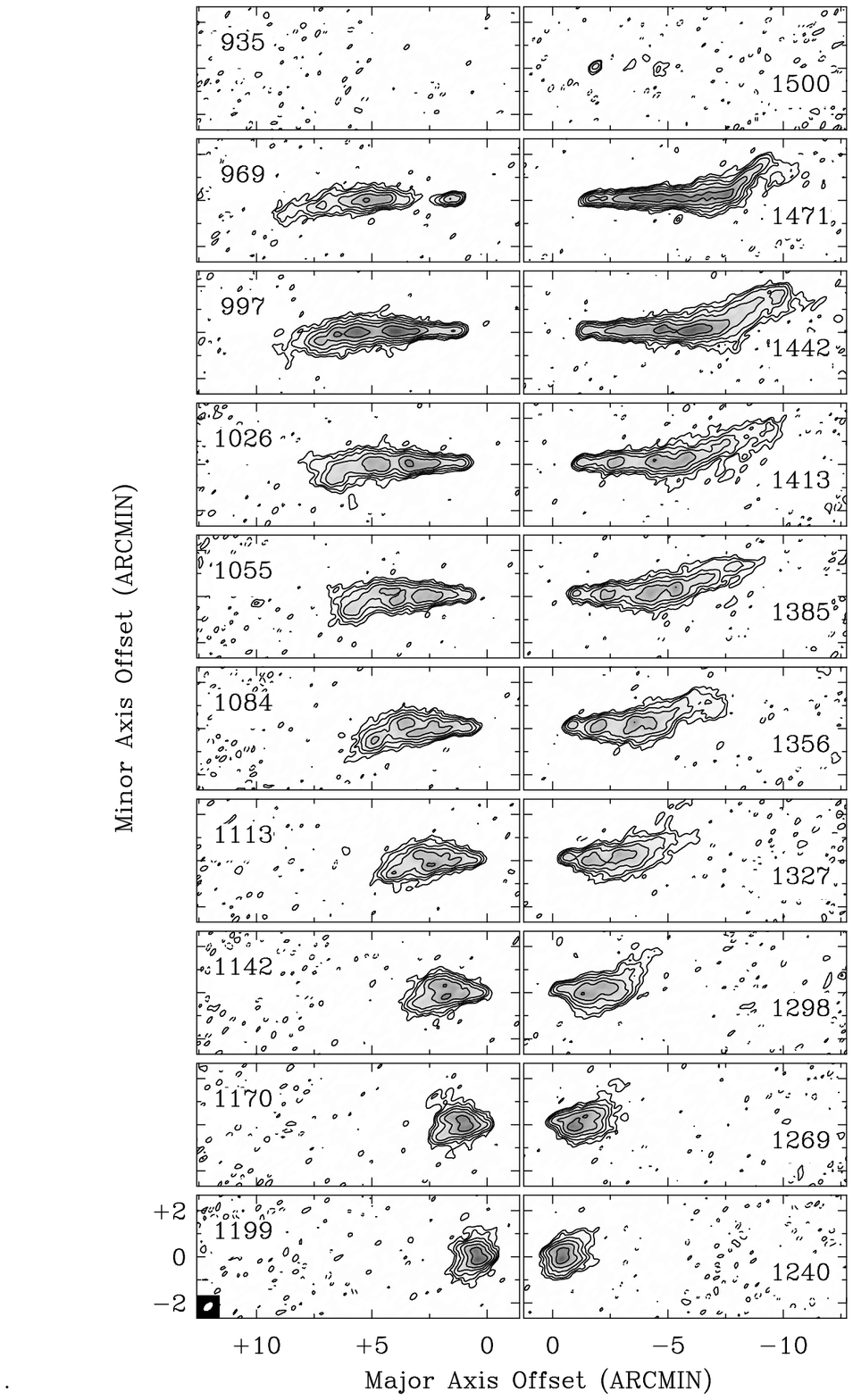}
\caption{Representative channel maps of the data with a resolution of 24.3"$\times$11.5".  Contours begin at 2$\sigma$ (0.61 mJy bm$^{-1}$) and increase by factors of 2. The $-2\sigma$ contour is also shown with dashed lines. The velocities in km s$^{-1}$ are given in each panel and the velocity spacing is approximately 7 resolution elements.  Note the slanted outer edges present in both halves, which correspond to the projection of the warp perpendicular to the line of sight as well as the asymmetric gas distribution in each half.  The beam is shown in the bottom left panel.  \label{fig2}}
\end{figure}

    The warp across the line of sight is also clear in the channel maps shown in Figure~\ref{fig2}.  Also seen more clearly in the channel maps is a substantial asymmetry in the distribution of gas in the approaching and receding halves.

\par
    There exist two nearby companions within the field of view: IC 3571 near the northeast quadrant of the galaxy with a velocity range of 1225-1299 km s$^{-1}$ and NGC 4562 in the southwest with a velocity range of 1266-1427 km s$^{-1}$.

\par
\begin{figure*}
\epsscale{1}
\figurenum{3}
\plotone{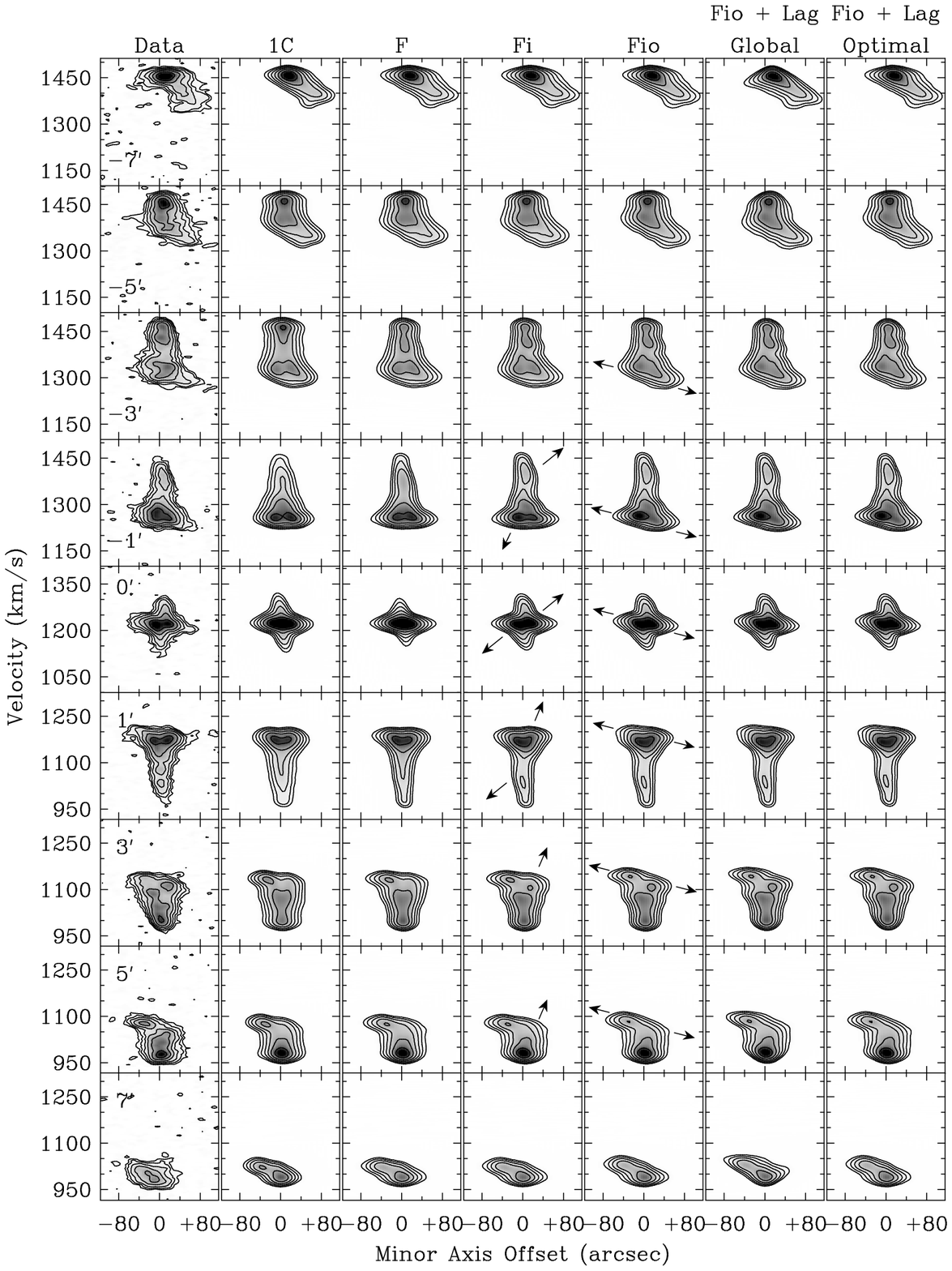}
\caption{Position-velocity diagrams parallel to the minor axis showing the data, one-component (1C) model, and various flaring (F) models. The Fi model is the flare model with disk inflow at small to moderate radii beyond the bar, and the Fio model is the flare model with both inflow in the inner parts of the disk and outflow in at large radii.  The second to last column shows the Fio model with a global lag of $-$10 km s$^{-1}$ kpc $^{-1}$, while the final column shows the Fio model with an optimized radially varying lag. Details for models involving radial motions are described in $\S$~\ref{Section4.1.5} and details of the lag models are described in $\S$~\ref{Section4.1.6}.  All models shown here include a bar with an orientation 30$\,^{\circ}$ from the line of sight and a radial motion of $-$20 km s$^{-1}$.  Contours are as in Figure~\ref{fig2}.  Slice locations are given in the first column and can be seen in Figure~\ref{fig1}. Arrows indicate the direction emission is moved due to the addition of radial motions.  The approaching half is SE while the positive minor axis offset is NE as seen in Figure~\ref{fig1}.\label{fig3}}
\end{figure*}

    There is a large velocity spread seen in minor axis position-velocity (bv) diagrams near the center of the galaxy [Figure~\ref{fig3} (column 1, row 5)].  As we will show in $\S$~\ref{Section4.1.3}, this velocity spread is kinematic evidence for a bar with an orientation that is at least partially along the line of sight with radial motions.

\par
    There also exist indications of radial motions \textit{not} associated with the bar.  These are best seen as a slanting of the data contours on the systemic side in most panels of Figure~\ref{fig3}.  The direction of the slant at minor axis offsets $<$ 40" is different from that at minor axis offsets 40" - 80".  Our modeling will show that this indicates a sign change in the direction of radial motions at larger radii rather than with height above the plane.

\par
      Crucial to certain aspects of the modeling, such as radial motions, is the distinction between the near and far side of NGC 4565.  Fortunately, from the orientation of the dust lane seen in the 2MASS Large Galaxy Atlas \citep{2003AJ....125..525J}, we can infer that the side northeast from the plane of the disk is nearer than the southwestern side.  This side corresponds to the positive side along the minor axis in the figures.

\par
     Modeling of the features listed above will be discussed further in the next section.

\par
\begin{figure}
\epsscale{1}
\figurenum{4}
\plotone{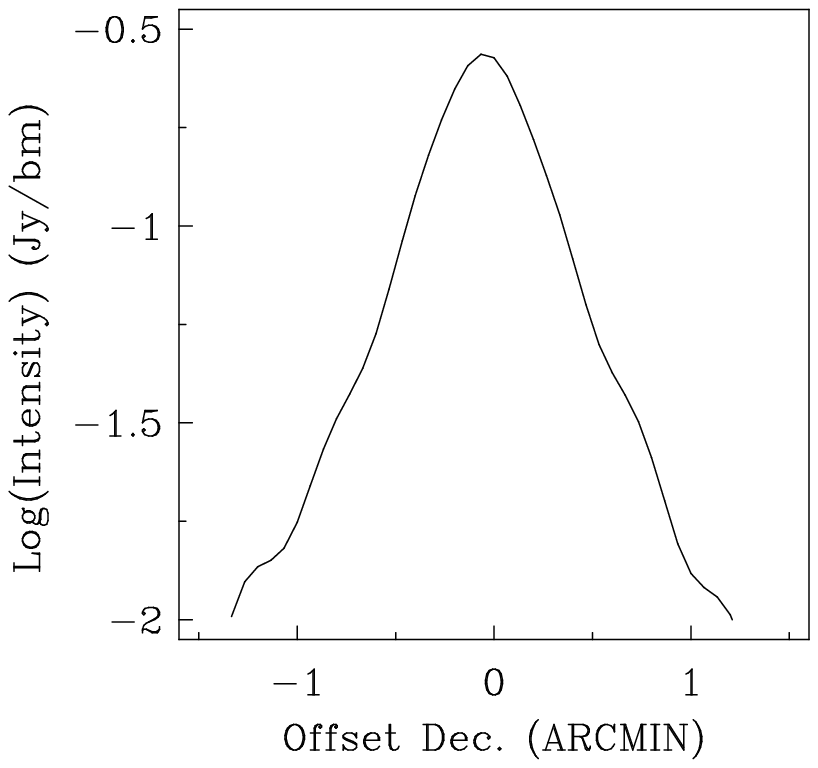}
\caption{The vertical profile of the {\sc H\,i} continuum at 30" resolution corresponding to major axis values between $-$4' and +4'.  The FWHM is approximately 1.2' or 3.8 kpc.  \label{fig4}}
\end{figure}

     Finally, from our continuum data we simply present a vertical profile in Figure~\ref{fig4}.  The detectable extent is comparable to that found by \citet{2008BASIP..25...67K}. 

\section{The Models} \label{Section4}

\par
   Models were created using the tilted ring fitting software ({\tt TiRiFiC}) \citep{2007A&A...468..731J}\footnotemark\footnotetext{http://www.astron.nl/{\raise.17ex\hbox{$\scriptstyle\sim$}}jozsa/tirific/}, which allows for a ${\chi}^{2}$ minimization fit.  Aside from its fitting capabilities, this software employs methods similar to the Groningen Image Processing System (GIPSY) \citep{1992ASPC...25..131V} task {\tt galmod}.  For each ring, parameters such as the central position, systemic velocity, inclination, position angle, surface brightness, rotational velocity, radial velocity, velocity dispersion and scale height may be specified. 

\par
   Contrary to models created using {\tt galmod}, those created using {\tt TiRiFiC} need not be axisymmetric.  {\tt TiRiFiC} allows the user to divide the model galaxy into wedges of azimuth in order to model asymmetries throughout the disk without requiring the creation of multiple models. Additionally, {\tt TiRiFiC} allows for a gradient in the rotational velocity with height.  {\tt TiRiFiC} can also superimpose gaussian components onto the disk, which can be used to resemble bars and spiral structure in the model.  Naturally, as is the case for tilted ring models in general, these are greatly oversimplified and should only be used as tools to approximate the morphology of these features and not to fully explain their kinematics.  

\par
   Although a majority of the modeling of NGC 4565 was done in {\tt TiRiFiC}, the earliest modeling steps were done using standard GIPSY tasks.  This is only because {\tt TiRiFiC} was still undergoing substantial improvements at the beginning of the modeling process. 

\par
\begin{figure}
\epsscale{1}
\figurenum{5}
\plotone{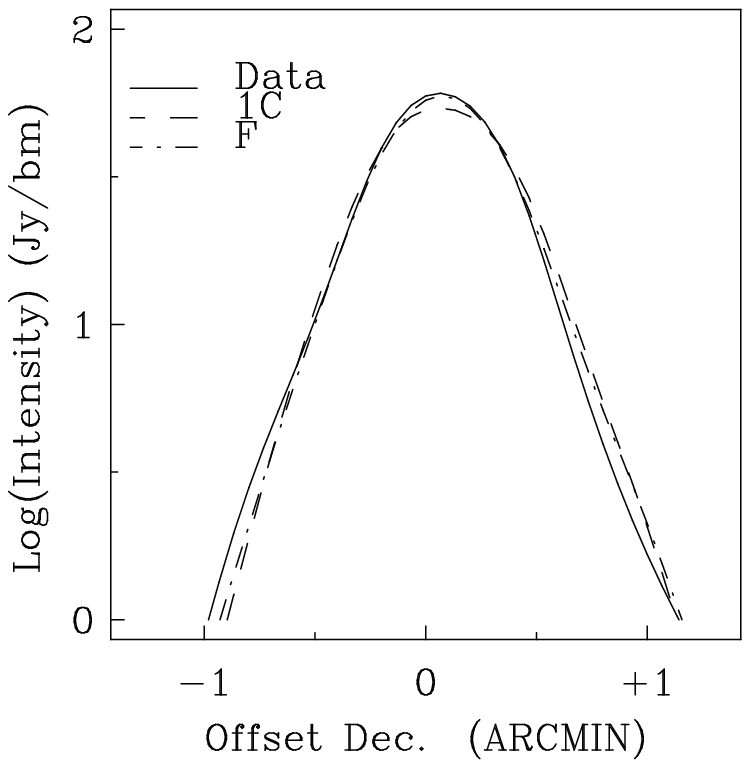}
\caption{The vertical profile of the data, one-component (1C), and flaring (F) models corresponding to major axis values between $-$4' and +4'.  Especially note the fit of each model near the peak of the profile as well as near the wings.  \label{fig5}}
\end{figure}

   An initial estimate for the systemic velocity was taken from \citet{1991AJ....102...48R}.  The rotation curves were determined via inspection of the terminal side of major axis position-velocity (lv) diagrams, akin to the envelope tracing method (\citealt{1979A&A....74...73S}, \citealt{2001ARA&A..39..137S}).  Surface brightness estimates were determined using the GIPSY task {\tt radial}.  The run of the position angle was initially estimated by eye and inclinations close to 90$\,^{\circ}$ were considered.  Initial estimates for the exponential scale height were determined solely by a fit to the vertical profile (Figure~\ref{fig5}).  All of these parameters were later refined both by comparing observed and modeled position-velocity diagrams, channel maps, zeroth-moment maps and vertical profiles.

\par
    {\tt TiRiFiC's} automated fitting capabilities were used when appropriate. However, at the level of detailed analysis of lopsided and asymmetric disks, which may be described by including non-standard parameterizations, this possibility is useful only to a certain degree. Nevertheless, if provided with reasonable initial constraints, {\tt TiRiFiC} may expedite the fitting process.  The degree to which additional refinement by eye is necessary depends greatly on any distinct features present in the region of fitting.

\subsection {Individual Models} \label{Section4.1}     

\par
    Several basic models are initially considered.  These include a simple one-component (i.e.\ single disk with a constant exponential scale height and no warp or flare), flare, a substantial warp component along the line of sight, and a model with a second, thickened disk simulating an extended global halo.  The one-component and flare models are labeled as 1C and F in figures. 

\par
   Upon examination of the zeroth moment map (Figure~\ref{fig1}) as well as the vertical profile (Figure~\ref{fig5}), the latter two models (not shown) may be eliminated quickly due to a lack of thickening of the gas layer at small to intermediate radii that is a feature of these models as well as poor fits to position-velocity diagrams and channel maps. The observed vertical profile is almost entirely devoid of wing structure, making it difficult to fit while maintaining the correct curvature and width near the center with such models.  A substantial warp component along the line of sight places flux high above the midplane on the systemic side of the bv diagrams as is seen in the data.  However, it also creates indentations along the major axis at large radii in channel maps as well as on the systemic side of bv diagrams.  These are not seen in the data.  As for a halo model, adding a second thicker component not only increases the width of bv diagrams on the systemic side, but also increases the width closer to the terminal side.  Again, this is not seen in the data.  For these reasons, only the 1C and F models are considered further and described below. 

\par
    Although multiple types of models are considered, some parameters remain fixed.  These include the central position (12h 36m 20.63s, 25d 59m 16.8s), systemic velocity (1220 km s$^{-1}$), velocity dispersion (10 km s$^{-1}$), the run of the position angle and inclination, the surface brightness profile, and the rotation curve. All models include a bar and a small-scale feature that are discussed in $\S$~\ref{Section4.1.3} and ~\ref{Section4.1.4}.

\par
    Due to asymmetries, the disk is divided into two halves along the minor axis, and each half independently modeled in what follows. 

\subsubsection{One Component Model} \label{Section4.1.1}

\par
\begin{figure*}
\figurenum{6a}
\includegraphics[angle=270,scale=.9]{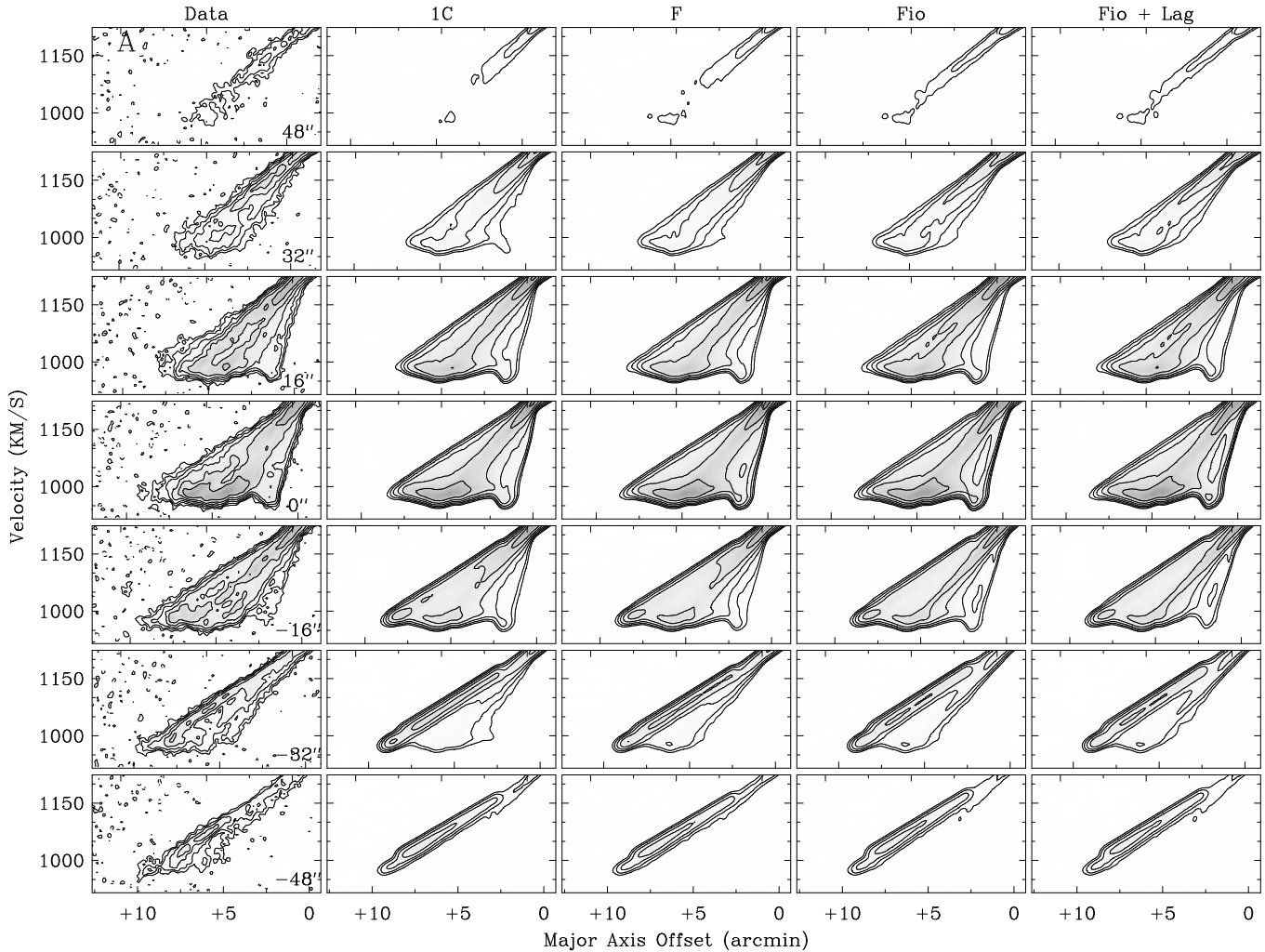}
\caption{Position-velocity diagrams of the approaching (A) and receding halves (B) parallel to the major axis showing the data, one-component (1C), and flaring (F) models.  The final two columns (Fio and Fio + Lag) display the flare model with radial motions and then with the optimal lag described in $\S$~\ref{Section4.1.5} and~\ref{Section4.1.6}.  All models shown here include a bar with an orientation 30$\,^{\circ}$ from the line of sight and a radial motion of $-$20 km s$^{-1}$. Contours are as in Figure~\ref{fig2}.  The heights above the plane of the disk are given in the first column.  Note that the lag is difficult to detect in these diagrams, primarily due to its being only within the inner 3-4'.  More convincing evidence is shown in Figure ~\ref{fig13}.  \label{fig6a}}
\end{figure*}

     A model consisting of a single component with an exponential scale height of 470 pc was created.  This model is a reasonable approximation to the data, but with deficiencies.  The first of these may be seen in the vertical profile (Figure~\ref{fig5}) where the 1C model is too flat near the top compared to the data.  If the scale height is decreased to match the data near the center, then the model becomes too narrow in the wings.  No other combination of scale height and inclination improves the fit. In the bv diagrams shown in Figure~\ref{fig3}, the 1C model is too thick at terminal and intermediate velocities when compared to the data.  Furthermore, an excess of emission in the 1C model can be seen at major axis offsets within $\pm$5' of the center offsets of the lv diagrams corresponding to  $-$16" and $-$32" in Figure~\ref{fig6a} and b. Finally, in Figure~\ref{fig7}, one may note that, while the minor axis thickness is well matched at velocities close to systemic, the 1C model is too thick at small and intermediate major axis offsets. As will be shown, these issues are all remedied by allowing the scale height to vary in the form of a flare.

\begin{figure*}
\figurenum{6b}
\includegraphics[angle=270,scale=.9]{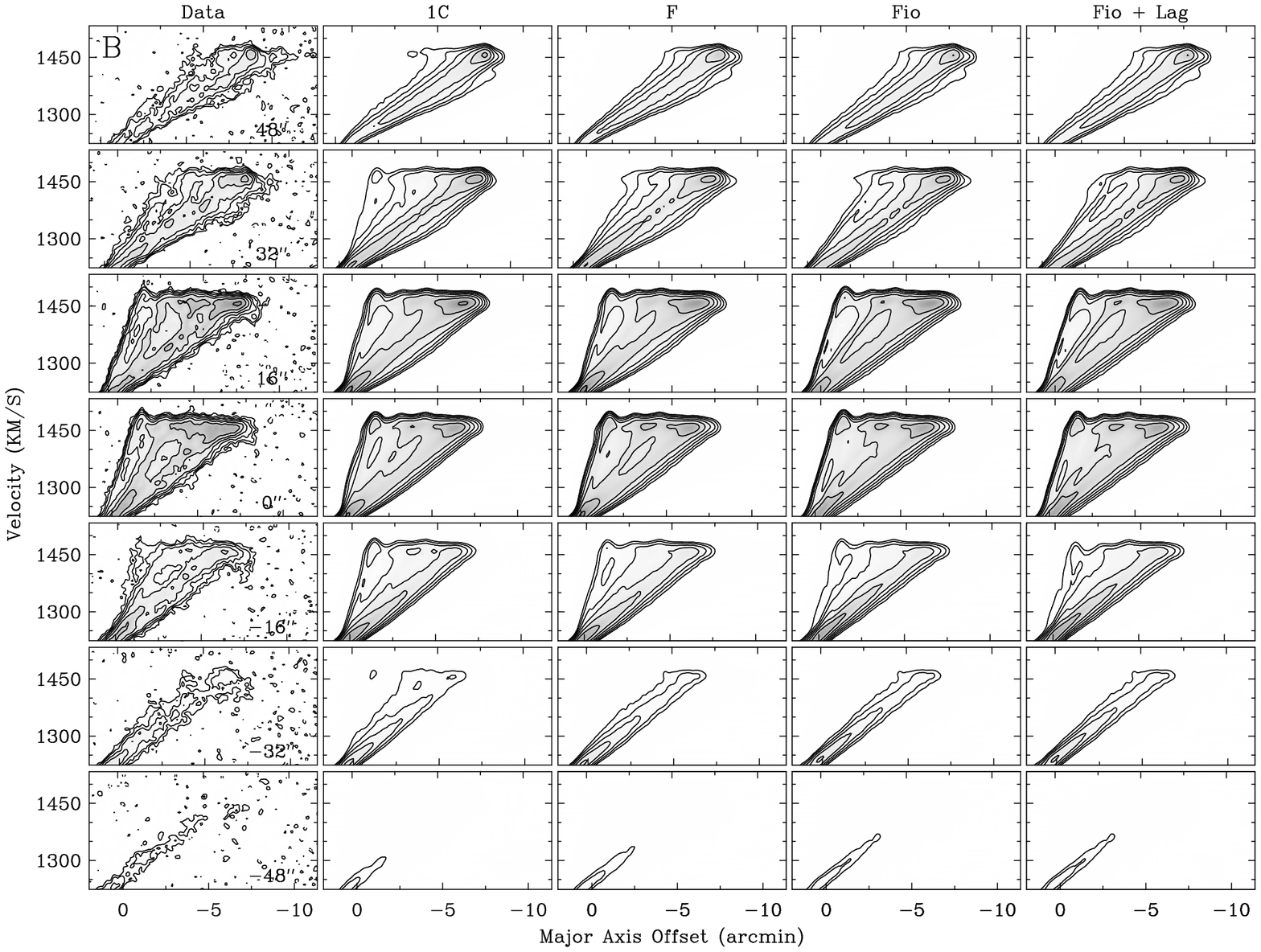}
\caption{\label{fig6b}}
\end{figure*}

\begin{figure*}
\figurenum{7}
\includegraphics[angle=270,scale=.9]{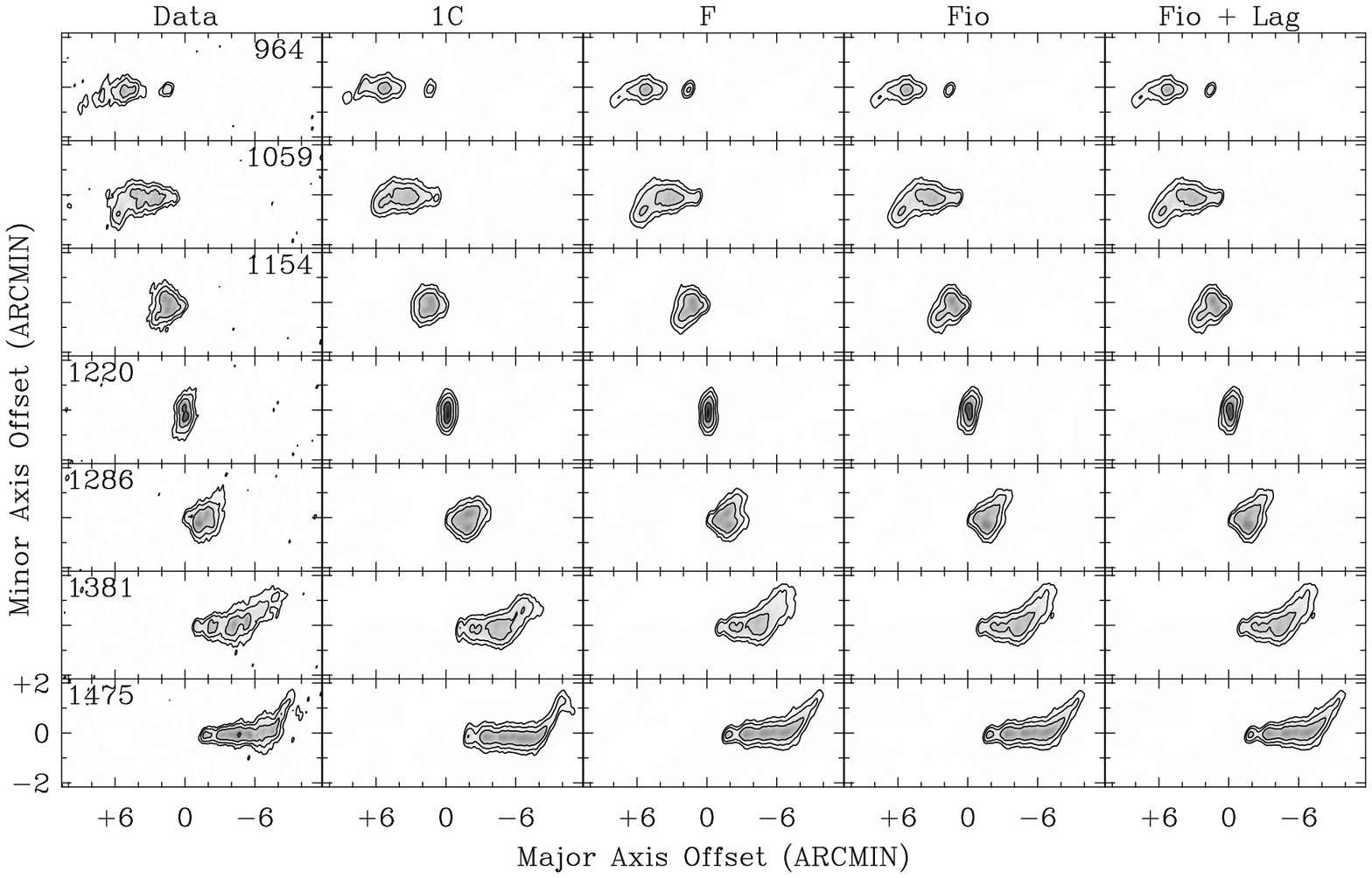}
\caption{Channel maps showing the data, one-component (1C) model, and various flaring (F) models. The final two columns (Fio and Fio + Lag) display the flare model with radial motions and then with the optimal lag described in $\S$~\ref{Section4.1.5} and~\ref{Section4.1.6}. Only subtle differences are seen with the addition of a lag, all within 3-4' from the center, most notably in the bottom row.  More convincing evidence is shown in Figure ~\ref{fig13}. All models shown here include a bar with an orientation 30$\,^{\circ}$ from the line of sight and a radial motion of $-$20 km s$^{-1}$.   Contours are as in Figure~\ref{fig2}.  Velocities are given in each panel of the first column. \label{fig7}}
\end{figure*}

\subsubsection{Flaring Model} \label{Section4.1.2}

\par
\begin{figure}
\epsscale{1}
\figurenum{8}
\plotone{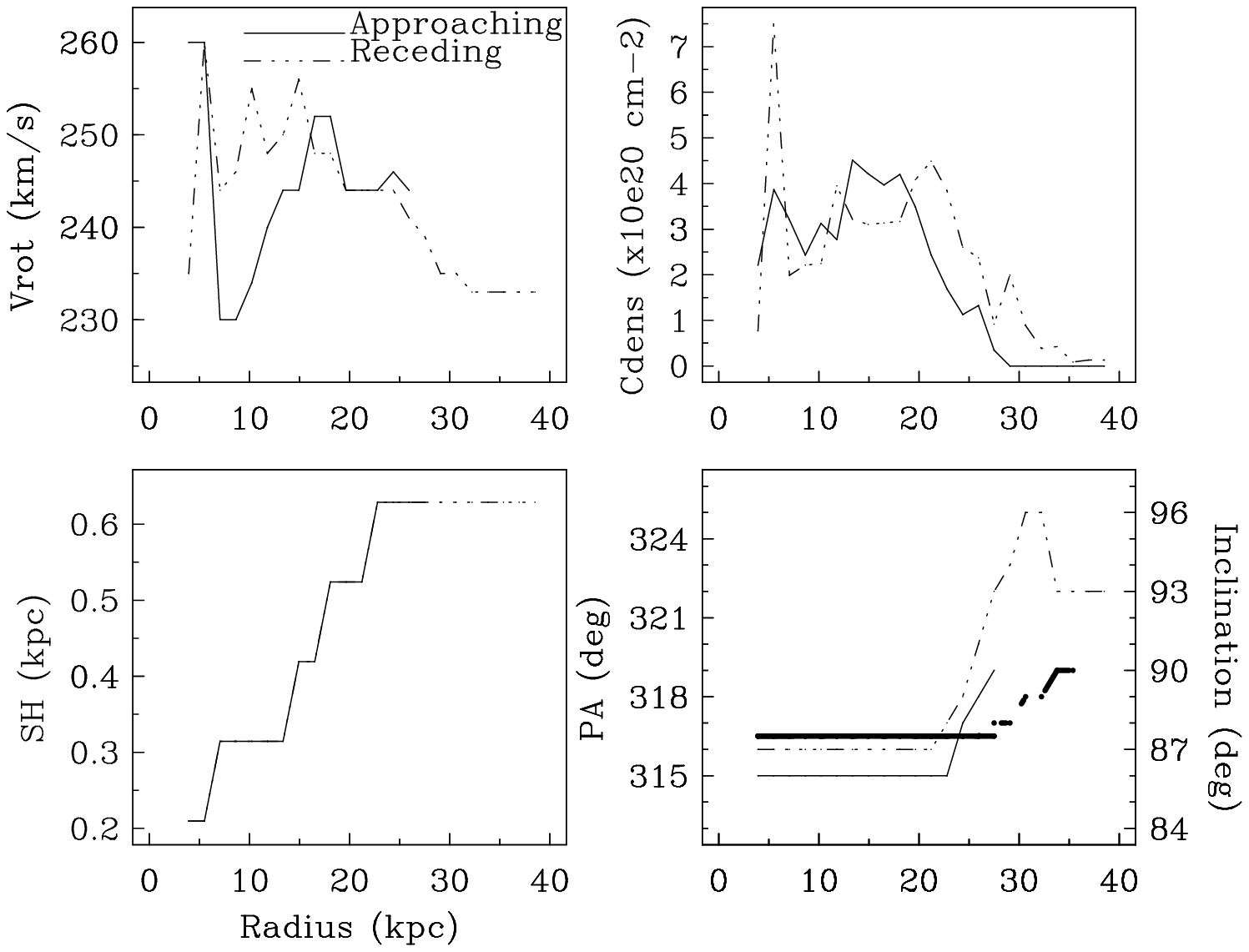}
\caption{Key parameters used in the optimal models, with the 1C and F models differing only in scale height.  These are rotational velocity (vrot), column density (Cdens), position angle (PA), inclination, and for the F model, scale height (SH).  Inclination is shown using thickened lines.  Note that the receding half extends further than the approaching half, which drops to zero in column density at approximately 30 kpc.  Values for other parameters beyond this point are excluded in that half.  \label{fig8}}
\end{figure}

     Instead of using a single exponential scale height throughout the disk, if we allow the disk to flare [scale height = 4" (~200 pc) near the center, increasing to scale height = 12" (~600 pc) in outer radii], the issues mentioned above are all but eliminated, as may be seen in the third columns of Figures~\ref{fig6a} and b.  Other variations in the scale height were tested, but this was found to be the best fit to the data.  Figure~\ref{fig8} shows the radial variation of scale height used in our flaring models.  Note that the scale height increases in steps rather than a straight line.  A smoothed version would produce nearly identical results, but arbitrarily smoothing the scale height distribution could imply that the models are better constrained than they are in reality.  Thus, we include the steps as they are a better representation of the modeling process and limitations. Nevertheless, the scale height is seen to rise in a nearly linear fashion.

\par
     Although some improvements with the flare model are seen in lv diagrams where the excess emission in the 1C model is not seen, the most convincing evidence for improvement is also seen in bv diagrams and channel maps (Figures~\ref{fig3} and~\ref{fig7}).  In Figure~\ref{fig3}, one may note a widening on the systemic relative to the terminal side of the data best seen in panels within $\pm$3', which is significantly better matched by adding a flare.  Additionally, in Figure~\ref{fig7} it is seen that there is a thickening at large major axis offsets in the channel maps that is best reproduced via the addition of the flare (e.g.\ the panel corresponding to 1381 km s$^{-1}$). 

\par
    Key parameters for the models described above, with the 1C and F models differing only in scale height behavior, are shown in Figure~\ref{fig8}.

\par
    Up to this point, we have discussed modeling only global features found throughout the disk.  Now we will discuss the modeling of more localized features such as a bar and adding further refinements such as radial motions.  Given that Figures~\ref{fig3} through 5 provide the most detailed representation of the models, these figures all include the features that will be described below. Abbreviated representations of the models excluding them will be shown in order to show their effects.

\subsubsection{Including a Bar} \label{Section4.1.3}

\par
\begin{figure}
\epsscale{1}
\figurenum{9}
\plotone{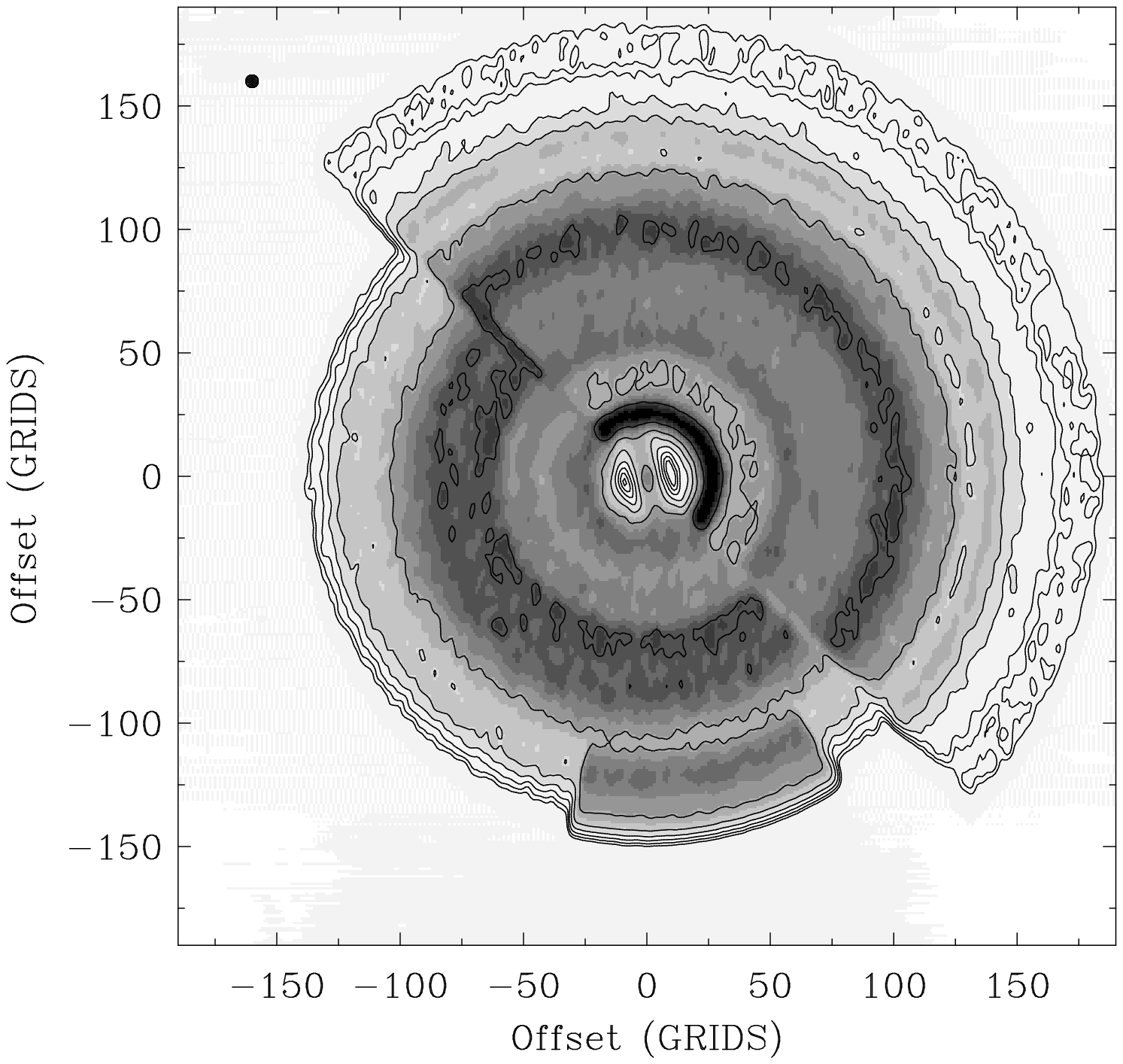}
\caption{A face-on view of the optimal model as seen from a viewer located in the SW quadrant.   Note the extent of each half, the bar at the center with a half-length of 1.25', the ring of higher density at approximately 1.7', as well as the additional arc added to the far side of the approaching half.  The black dot indicates the original orientation of the observer.  Contours begin at $7.7\times10^{15}\mathrm{cm^{-2}}$ and increase by factors of 2.  Grid units are in pixels and 1 pixel = 4". \label{fig9}}
\end{figure}
\begin{figure*}
\epsscale{1}
\figurenum{10}
\plotone{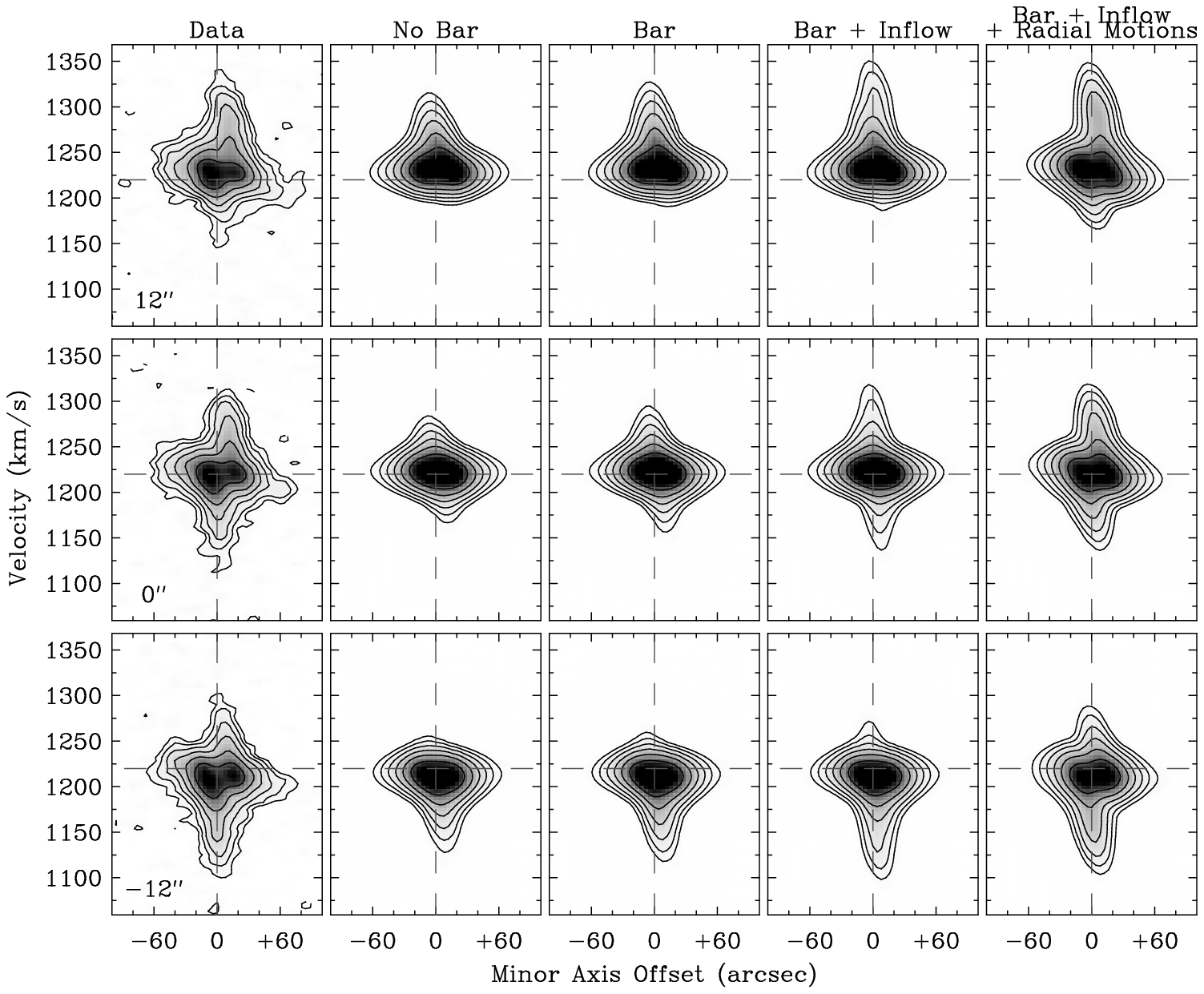}
\caption{Minor axis position-velocity diagrams showing models with no bar, a bar oriented 30$\,^{\circ}$ from the line of sight, the same bar with an inflow of $-$40 km s$^{-1}$, and the same bar with an inflow of $-$20 km s$^{-1}$ as well as additional radial motions that are not associated with the bar.  These additional radial motions are described in $\S$~\ref{Section4.1.5}.   Note the large spread in velocities in each panel related to the bar itself, as well as the slanting of the contours at 1175-1250 km s$^{-1}$, indicative of additional radial motions.  Contours are as in Figure~\ref{fig2}. Dashed lines indicate the major axis and systemic velocity.\label{fig10}}
\end{figure*}

    As mentioned in $\S$~\ref{Section1} and ~\ref{Section3} there are indications of a bar in NGC 4565, which we will show can also explain the large velocity spread in panels near the center in Figures~\ref{fig3} and ~\ref{fig10}.  The effects of adding a bar are most clearly seen in the latter Figure, which shows no-bar and bar models side-by-side.  We model this bar by adding a series of 1D gaussian emission components starting at the center and moving outward radially along some azimuthal direction in the disk.  The dispersion of the gaussian component is specified along the azimuthal angle and is 15" in our models.  The amplitude is also specified.  Additionally, parameters which are specified in rings as described above may also be set, notably the azimuthal position and radial velocity.  It should be noted that these parameters allow for an approximation of a bar, and do not account for more complex streaming motions or morphology.  Figure~\ref{fig9} shows how this bar appears in a face-on view of our best model. 

\par
     We examine several orientations with respect to the line of sight, keeping in mind that \citet{2010ApJ...715L.176K} assume a line of sight (i.e.\ end-on) orientation due to the boxy bulge seen in the optical, while \citet{2005MNRAS.358.1477A} states that such boxy bulges are indicative of components across the line of sight (i.e.\ side-on), but possibly only as little as 10$\,^{\circ}$ from the end-on view. From the clear indications of radial motions described below, we can be certain that the bar is at least partially end-on, although it is difficult to gauge how much. We explore this in our models.

\par
   First we examine a line of sight or end-on orientation and assume that the half-length of the bar roughly extends to 1.25' (4 kpc), or just within a ring of higher density in the disk (Figure~\ref{fig8}).  A linearly increasing rotation curve is used, which matches the rest of the disk at radii larger than 45".  (Note:  aside from the bar, no {\sc H\,i} is included in the main disk interior to this radius.)  Adding a bar in this way allows for some spread in the velocity range (Figure~\ref{fig10}, column 3), while remaining in a narrow range along the minor axis.  However, the full velocity spread seen in the data is not achieved. Such a model is improved by adding radial motions of $\pm$20 km s$^{-1}$ (Figure~\ref{fig10}, column 4). Due to projection effects and limited spatial resolution, we cannot discern whether this is inflow or outflow.

\par
    For $\pm$30$\,^{\circ}$ and $\pm$45$\,^{\circ}$ offsets from the line of sight (positive corresponds to the near side of the bar being in the NW quadrant), we achieve a somewhat improved fit with radial motions of $\pm$40 km s$^{-1}$.  Finally, increasing the offset to $\pm$60$\,^{\circ}$ from the line of sight requires increasing the radial motions to $\pm$60 km s$^{-1}$.  These results ignore any additional radial motions in the disk beyond the bar, which are shown to be necessary in $\S$~\ref{Section4.1.5}.  These motions in the disk render constraining the orientation and radial motions of the bar even more difficult.

\par
     The length of the bar and the dispersion of the gaussian distortions remain constant in all cases.  For the figures in this paper we display a $-$30$\,^{\circ}$ offset bar with a radial motion of $-$20 km s$^{-1}$.  It should be noted that +30$\,^{\circ}$ paired with +20 km s$^{-1}$ produces the same results. The disk radial motions decrease the amplitude of the optimal bar radial motion, hence we choose $-$20 km s$^{-1}$ for the latter instead of $-$40 km s$^{-1}$ as mentioned above.

\subsubsection{Additional Morphological Modifications} \label{Section4.1.4}

\par
     In the approaching half, a clump of brighter emission is seen between $-$20 and $-$50" above the plane of the disk, centered at a velocity of 1075 km s$^{-1}$ (Figure~\ref{fig3}, panel corresponding to 5').  This is not to be confused with the slant due to radial motions mentioned in $\S$~\ref{Section3}, and described in $\S$~\ref{Section4.1.5}.  To sufficiently achieve both of these elements, an arc centered at $-$55$\,^{\circ}$ from the line of sight in azimuth within the disk, and spanning 45$\,^{\circ}$, is added to the far quadrant of the approaching half in all models (Figure ~\ref{fig9}). This is likely an azimuthal continuation of the larger radial extent of the receding half relative to the approaching half, or possibly spiral structure, rather than a unique feature.  The modeling of such features as spiral structure will be explored in future HALOGAS papers.

\subsubsection{Radial Motions in the Disk} \label{Section4.1.5}

\par
    There are several indications of radial motions beyond the bar.  This is most readily seen in Figure~\ref{fig3} as well as in Figure~\ref{fig10}.  Note first the slope of the highest contours on the systemic side of the approaching half, particularly evident in the panels corresponding to 0' and 1'.  This is appropriately reversed for the receding half.  These are indicative of radial inflow at small to intermediate radii.  This is modeled by introducing an inflow starting with an extreme value of $-$50$\pm$5 km s$^{-1}$ at 1.75' (5.5 kpc) and quickly decreasing with radius before becoming undetectable beyond 2.75' (8.6 kpc).  This inflow is symmetric in both halves and is very likely associated with the ring of higher density. The model in the fourth column of Figure~\ref{fig3} (Fi) includes all of these motions. Arrows in Figure~\ref{fig3} indicate the significant but often subtle effects of adding such inflow.  The directions of the arrows correspond to the direction in which emission is displaced due to radial motions.  Additional isolated regions of inflow are seen in each half, but these are not symmetric and are likely associated with spiral structure.  It should be noted that a more modest inflow of 10-15 km s$^{-1}$ spread over a larger radial range could not reproduce what is seen in the data.  

\par
     In addition to the indications of radial inflow in the inner parts of the galaxy, indications of outflow are seen at large radii.  This is best seen as the overall negative slope on the systemic side, opposite in direction from that produced by the inflow, $\pm$40-80" off the plane of the disk in panels between -5' to +5' (once again indicated by arrows).  This is modeled by allowing a radial outflow of 10 km s$^{-1}$ beginning at 6.75' (21.2 kpc), and extending outward to the edge of the disk in each half [Figure~\ref{fig3}, column 5 (Fio)]. 

\par
\begin{figure}
\epsscale{1}
\figurenum{11}
\plotone{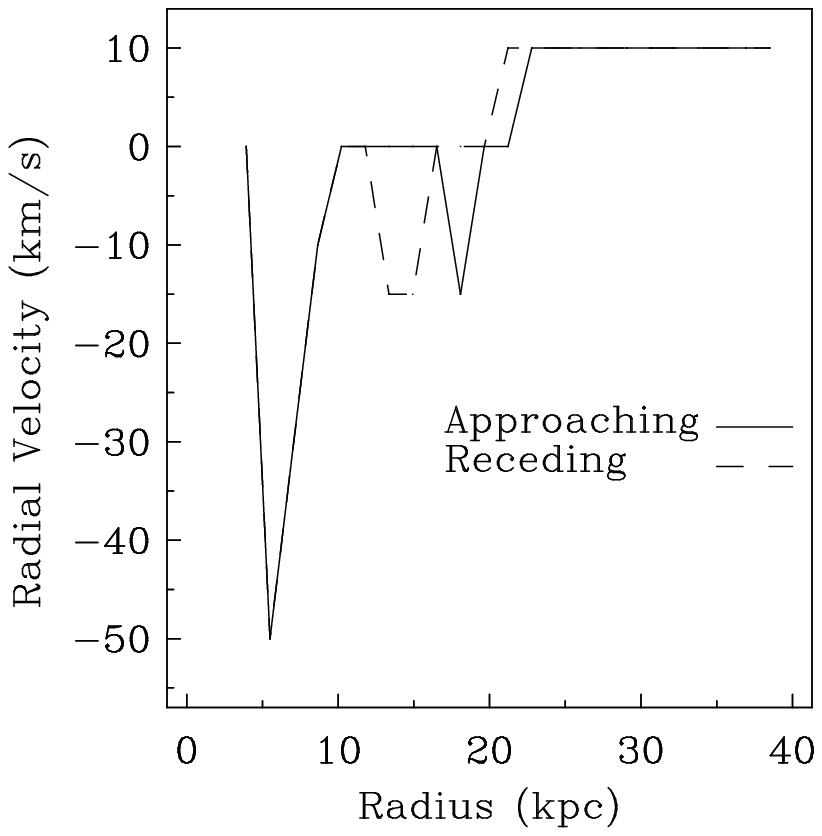}
\caption{Radial motions in each half, excluding those associated with the bar.  Negative values indicate inflows.  Due to limited resolution, and complications from the bar, no radial motions are constrained for radii less than 45" (2.4 kpc).  The sharp dip coincides with the ring of higher density, while the shallower dips are likely associated with spiral structure.\label{fig11}}
\end{figure}

     The distribution of radial motions in the models may be seen in Figure~\ref{fig11}.

\par
     It is difficult to tell with certainty if the observed radial motions are due to streaming along spiral arms or another source.  However, the substantial inflow located within and just outside the ring of higher density indicates that the inflow in this region may be true axisymmetric inflow or associated with the bar rather than spiral arms.  

\subsubsection{The Addition of a Lag} \label{Section4.1.6}

\par
\begin{figure*}
\epsscale{1}
\figurenum{12a}
\plotone{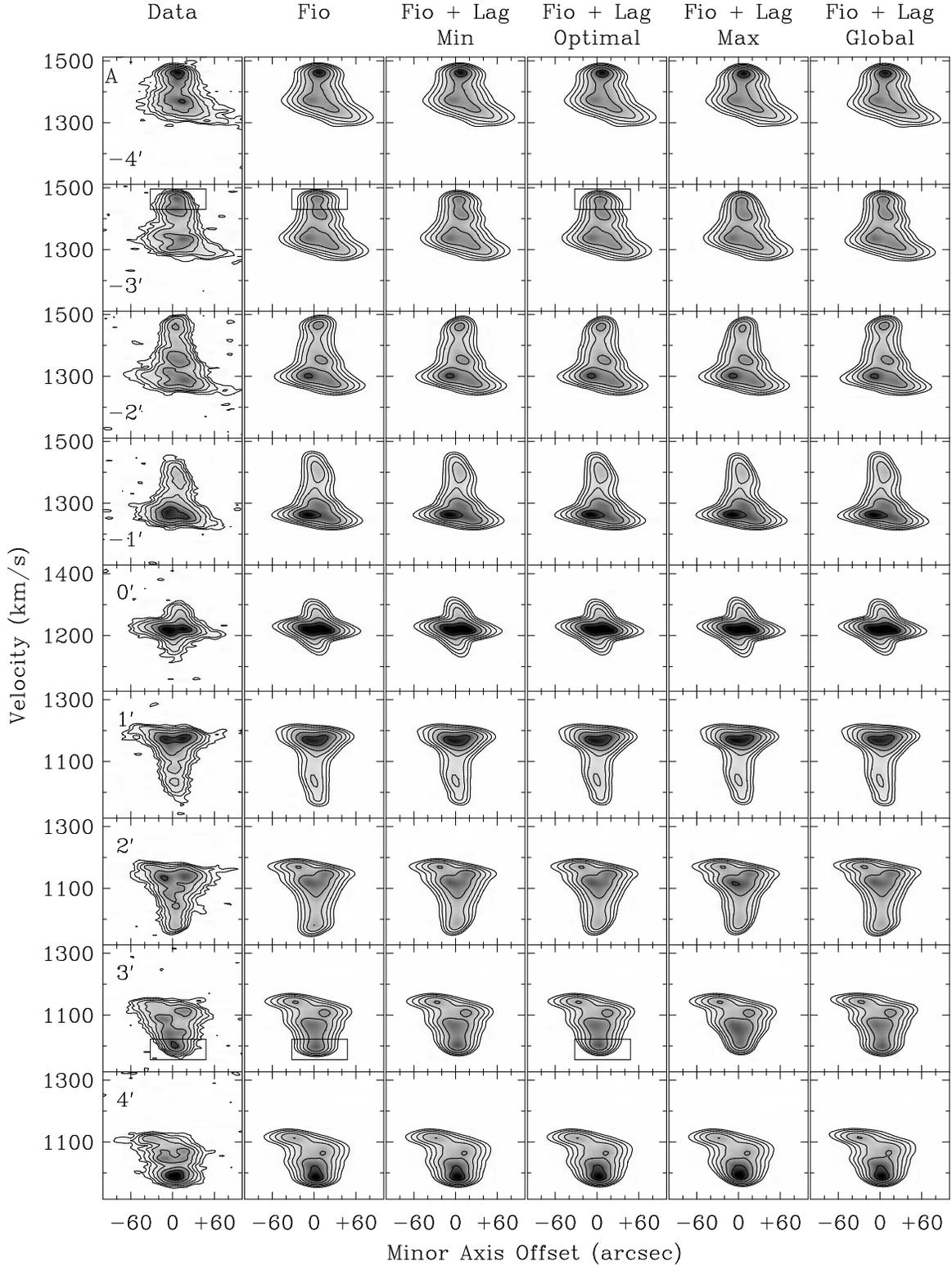}
\caption{Position-velocity diagrams parallel to the minor axis showing the data, flaring (Fio), and several models constraining different lags, the parameters of which are given in Figure~\ref{fig13}.  The optimal lag is included in the Fio + Lag model, while FLmin and FLmax are models with the lower and upper limits.  The final column shows a global lag of $-$10 km s$^{-1}$ kpc$^{-1}$ that fits the data well near the center, but as can be seen in Figure~\ref{fig3} does not match the data at large radii. Contours are as in Figure~\ref{fig2} and slice locations are given in the first column.  Rectangles indicate regions where effects of the lag are most noticeable, and the panels containing them are magnified in Figure~\ref{fig12b}. \label{fig12a}}
\end{figure*}

     As previously discussed in $\S$~\ref{Section4.1.1} and seen in the bv diagrams in Figure~\ref{fig3}, the data are relatively thin near the terminal side compared to the systemic.  This is somewhat remedied by the addition of a flare ($\S$~\ref{Section4.1.2}), but improvements may still be made through adding changes in rotational velocity with height above the disk.  This causes observed velocities of emission at high z to be displaced towards the systemic side in the bv diagrams, resulting in a thinning in the width near the terminal side and a thickening on the systemic side.  This is best seen in the bv diagrams closer to the center, shown in Figures~\ref{fig12a} and b where the contours on the terminal side become more rounded to better match the data with the addition of a lag. This is also evident at the terminal edge of lv diagrams, where emission is displaced to lower velocities.  It should be noted that these effects are subtle but cannot be reproduced by changing the flare parameters. If a lag is present in the data, then we should see some or all of these features, provided there is sufficient resolution.  However, given how thin and far away the disk is, our resolution is lacking, so our error bars are large.   We find an optimal lag peaking at -40 $^{+5}_{-20}$ km s$^{-1}$ kpc$^{-1}$  between 1.25' and 4.75' (3.9 and 14.9 kpc) in the approaching half, and -30 $^{+5}_{-30}$  km s$^{-1}$ kpc$^{-1}$ between 1.25' and 4.25' (3.9 and 13.4 kpc) in the receding half.  These quickly decrease in magnitude within the specified range in both halves, as shown in Figure~\ref{fig13}.  This lag distribution is again represented in Figure~\ref{fig14} as an azimuthal velocity  that rises more steeply at high z. 

\begin{figure}
\epsscale{1}
\figurenum{12b}
\plotone{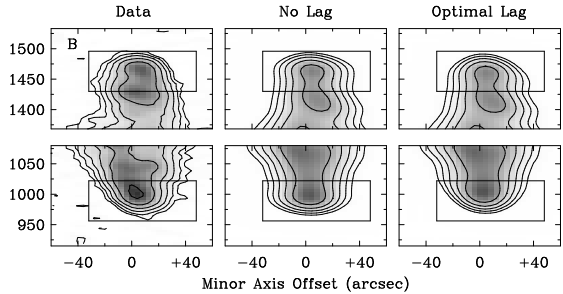}
\caption{Magnified panels corresponding to those with rectangles in Figure~\ref{fig12a}.  Note the rounding and slope of the contours near the terminal side in each panel.\label{fig12b}}
\end{figure}

\begin{figure}
\epsscale{1}
\figurenum{13}
\plotone{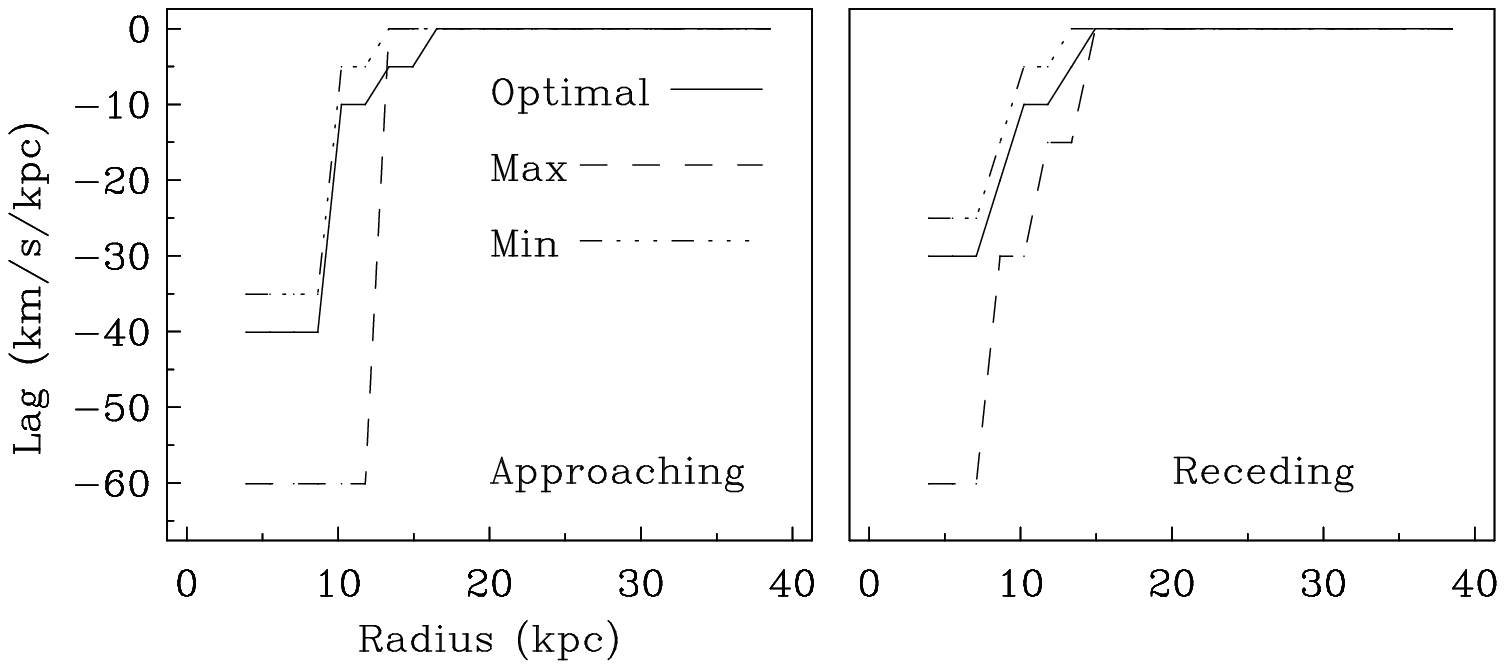}
\caption{The optimal, maximum and minimum lag distributions in each half.  Due to limited spatial resolution, no lag is constrained for radii less than 45" (2.4 kpc).\label{fig13}}
\end{figure}

\par
     For completeness we also include a model with a constant global lag throughout the galaxy in the sixth column of Figure~\ref{fig3}.  Such a lag overestimates the rounding on the terminal side in panels corresponding to $\pm$5' and $\pm$7', emphasizing the need for radial variation in the lag.

\begin{figure}
\epsscale{1}
\figurenum{14}
\plotone{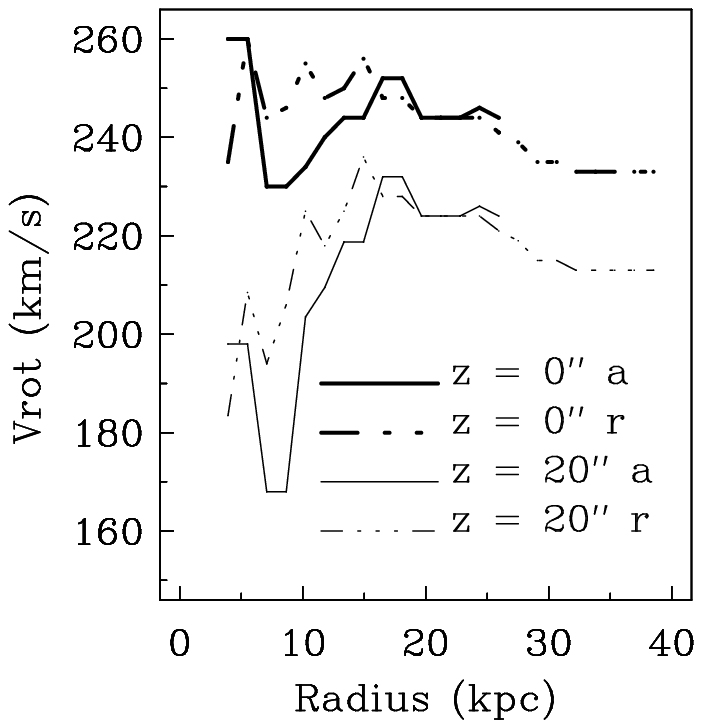}
\caption{Azimuthal velocities in the midplane and 20" ({\raise.17ex\hbox{$\scriptstyle\sim$}}1 kpc) above the plane of the disk for the approaching (a) and receding (r) halves.\label{fig14}}
\end{figure}

\subsection{On the Uniqueness of the Models} \label{Section4.2}

\par
     A rigorous statistical analysis of the fitting of tilted ring models is generally omitted from the literature (e.g.\ \citealt{2005A&A...439..947B}, \citealt{2007AJ....134.1019O},
\citealt{2008ApJ...676..991R}).  Such studies employ a simple model focusing on the quantification of certain features in observed galactic disks, driven by a specific scientific question. In our case
the main aim is to constrain the vertical structure of the {\sc H\,i} disk, which means giving significant weight to faint emission.  If fitting were fully automated, a goodness-of-fit criterion such as a ${\chi}^{2}$ (which {\tt TiRiFiC} provides) could be used to distinguish models by significantly downweighting the bright emission that would otherwise dominate the statistic and render it meaningless for our purposes.  In principle, a reduced ${\chi}^{2}$ with such a weighting could also account for the varying number of free parameters among models.  However, it is generally much more efficient to use the judgment of the modeler and adjust certain parameters manually.  But this makes an accurate accounting of the degrees of freedom in each model difficult.  For instance, the radial scale height dependence (Figure~\ref{fig8}), shows eight values and eight inflections, indicating sixteen values were varied.  However, for this and other parameters, finer variations may have been considered and discarded as providing insignificant improvement.  Likewise, initially most parameters for the approaching and receding models are the same, and then some parameters varied as needed, but some deviations between halves may have been tested and discarded to keep the models simple.  If so, the number of parameters allowed to vary was larger than the final model would suggest.

\par
    We therefore rely on fitting by eye to guide our choice of models.  A visual comparison of data and models provides a powerful method for deciding how models must be altered to fit the data, easily allowing attention to be focused on regions of interest.  One statistic we could examine, regardless of the number of degrees of freedom, is the rms of the residual cube, downweighting regions of bright emission.  In fact, when we examine this statistic, using only emission above 3$\sigma$, we indeed find a significant 7$\%$ reduction for models which include a flare (all models beginning with ``F") rather than a constant scale height (1C).  However, among the flaring models, the numerical difference due to the addition of radial motions, a bar, or a lag is negligible.  Nevertheless, improvement is clearly seen in Figures~\ref{fig3}, ~\ref{fig10}, ~\ref{fig12b} respectively.  In spite of such limitations, we address the need for such features here.

\par 
    Firstly, the additional arc in the receding half ($\S$~\ref{Section4.1.4}), which is included in all models presented in this paper, can be justified rather simply.  There are clear indications that a feature of higher flux density exists in this region, but only in one quadrant of the disk (Figure~\ref{fig3}, panel corresponding to 5').  The feature could be reproduced with an overall change in position angle, allowing flux to be redistributed at high z, but as this feature is seen only in a select region of the disk, such a change would worsen the fit in other parts of the disk.  Additionally, a change in position angle for only this region would be unphysical and was seen to be a less desirable match during the modeling process (not shown).  

\par
    The addition of radial motions duplicates slants, slopes, and curves in the data (Figure~\ref{fig3}) that are otherwise not reproducible.  One could argue for changes in the position angle near the center to reproduce these slants, but through modeling, we see that such a change would not fit the data as well.  Furthermore, warping near the center, as well as both the positive and negative changes in position angle required to reproduce the correct slanting would be uncharacteristic of warping typically seen in other galaxies \citep{1990ApJ...352...15B}.

\par
    It is clear that a bar exists in NGC 4565, both from previous works at other wavelengths (\citealt{2010ApJ...715L.176K}, \citealt{1996A&A...310..725N}), as well as our models.  Substantial improvement can be seen in Figure~\ref{fig10}, although due to projection effects, the uniqueness of this bar is difficult to constrain. Thus we provide generous uncertainties for the quantities involved ($\S$~\ref{Section4.1.3}).

\par
    Finally, the lag is best constrained using faint emission, high above the plane of the disk.  For this reason, improvements due to a lag do not exhibit substantial changes in the rms of the residuals.  Nonetheless, subtle improvements are seen in Figure~\ref{fig6a} and b, as well as Figure~\ref{fig12a} and b.  The characteristics of the lag may initially be interpreted as {\sc H\,i} that appears (at least in projection) above the plane of the disk that is rotating more slowly than {\sc H\,i} in the midplane.  The issue of projection effects \textit{could} hinder the uniqueness of our models if such effects are not considered carefully.  However, signatures in both bv diagrams and channel maps described at the beginning of $\S$~\ref{Section4.1} allow us to successfully rule out inclination effects in this case.  From this we may conclude that the observed effects are extremely likely due to a lag.  A range of possible values for this lag is given in Figure~\ref{fig13}.

\section{The Companion Galaxies} \label{Section5}
\par
  While NGC 4562 is not our primary target, some information may be extracted from a basic model.  We assume a distance of 12.5 Mpc based on values obtained from the NASA/IPAC Extragalactic Database (NED).  The total {\sc H\,i} mass for this galaxy (including a factor of 1.36 to correct for He) is 3.5$\times$10$^{8}$ $M_{\odot}$.  Our models include flux extending out to a radius of 2.75' (10 kpc).   This small angular size limits our modeling capabilities in this case.  The scale height is found to be about 7" (~400 pc).  The galaxy is slightly inclined at 82$\,^{\circ}$.  There are no indications of a strong warp component along the line of sight, although there exists a slight warp component across it.  The rotation curve is relatively flat, and peaks at 65 km s$^{-1}$. 

\par
    No model is created for IC 3571.  However, assuming approximately the same distance as NGC 4565, its {\sc H\,i} mass is found to be 5.6$\times$10$^{7}$ $M_{\odot}$ and the extent of its largest axis is approximately 1.8' (5.8 kpc).

\section{Discussion} \label{Section6}

\subsection{The Warp in NGC 4565 and Interactions With Companions} \label{Section6.1}

\par
     The warp in NGC 4565 is highly asymmetric, not only extending further and reaching higher values in position angle on the receding half, but also flattening at large radii on that side as well (Figure~\ref{fig8}).  There was a hint of this flattening in a few channels as noted by \citet{1991AJ....102...48R}, most prominent in the channel corresponding to $v_{hel}$ = 1459 km s$^{-1}$.  Here we see it from $v_{hel}$ = 1440 to 1473 km s$^{-1}$.  A similar flattening is also seen in NGC 5907 \citep{1983IAUS..100...55S}.  

\par
     Additionally, there is a slight rise in inclination for radii larger than 8.75' (27.5 kpc), causing the galaxy to become more edge-on.  This is only seen in radii beyond the extent of the emission in the approaching half, thus it is only seen in the receding half.

\par
    It is difficult to determine the exact cause of the asymmetries in the warp, but it is possible that the companion galaxies play some role.  NGC 4565 is clearly interacting with IC 3571, as can be seen in Figure~\ref{fig1} by the bridge of material between them.  Additionally, material closer to the plane of the disk appears to be displaced above the plane, toward IC 3571 (Figure~\ref{fig1}).  Given the location of IC 3571, it may contribute to the asymmetry in the component of the warp across the line of sight. This was also noted by \citet{2005ASPC..331..139V}.

\par
\begin{figure}
\epsscale{1}
\figurenum{15}
\plotone{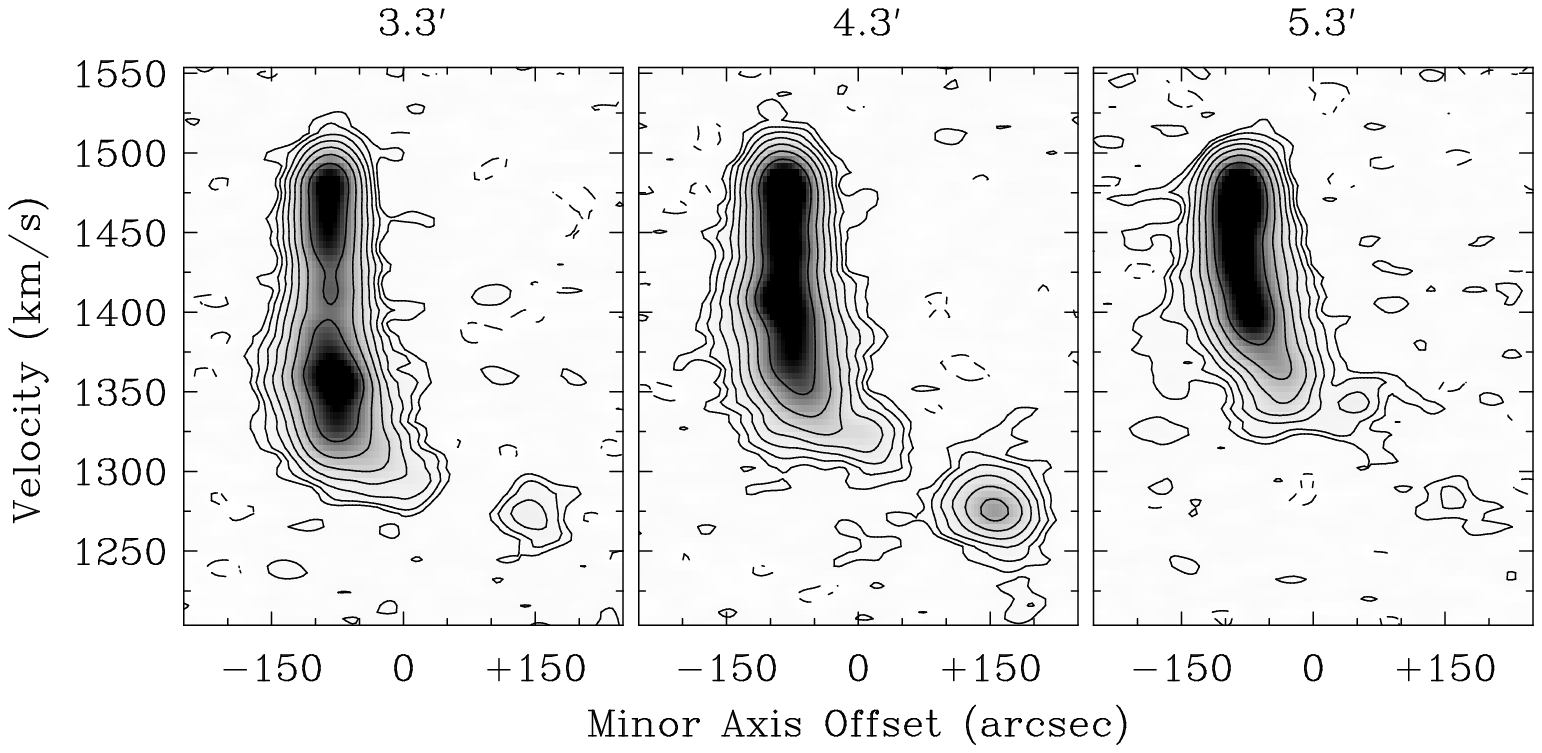}
\caption{Position-velocity diagrams along the minor axis showing the bridge between NGC 4565 and IC 3571 in the smoothed ($44"\times34"$) cube.   Contours begin at 2$\sigma$ (0.48 mJy bm$^{-1}$) and increase by factors of 2. A $-2\sigma$ contour is shown as dotted lines. Major axis offsets are given at the top of each panel.\label{fig15}}
\end{figure}

     Emission likely tied to IC 3571 as part of the bridge between the two galaxies could not be reproduced in models, which is unsurprising.  The velocity range associated with this emission is reasonably close to that of the main disk of NGC 4565 (Figure~\ref{fig15}), and is likely only a disturbance of gas within the disk as opposed to accretion from IC 3571.

\subsection{The Flaring Gas Layer} \label{Section6.2} 

\par
     The exponential scale height in NGC 4565 begins to increase in both halves from an initial 4" (~200 pc) near a radius of 3 arcminutes (9.4 kpc), eventually reaching a value of 12" (~600 pc) at a radius of 7 arcminutes (22.8 kpc).  This is just within the optical radius (8.1' or 25.5 kpc).   

\par
     Flares are seen in other galaxies, including M 31 \citep{1984A&A...141..195B} and the Milky Way (\citealt{2007A&A...469..511K} and references therein).  However, there exist galaxies for which extensive {\sc H\,i} modeling has been done that do not show flaring gas layers, including NGC 5746 \citep{2008ApJ...676..991R} and NGC 4559 \citep{2005A&A...439..947B}, indicating that flares are common although not ubiquitous.

\subsection{Halo Trends} \label{Section6.3}

\par
     Our models yield no evidence for a massive, extended {\sc H\,i} halo in NGC 4565.  This is shown by the vertical profile in Figure~\ref{fig5} where there are no indications of a second, extended component that exists globally throughout the data.  Furthermore, a second component is not seen to improve position-velocity diagrams or channel maps (not shown).  

\par
     We return to the issue discussed in $\S$~\ref{Section1} of whether there is any connection between {\sc H\,i} halos and disk star formation activity.  It is difficult to extract any concrete trends regarding {\sc H\,i} halos with such a small number of galaxies having been observed to this depth and subsequently modeled in detail.  The situation is in the process of being remedied via the HALOGAS survey and Expanded Very Large Array (EVLA) observations involving some HALOGAS team members.  In Table~\ref{tbl-3} we compare properties including total {\sc H\,i} mass, total infrared luminosity per unit area, scale heights, and lags of NGC 4565 with those of previously modeled edge-on galaxies.

\begin{deluxetable*}{lcccccc}
\tabletypesize{\scriptsize}
\setlength{\tabcolsep}{0.05in} 
\tabletypesize{\scriptsize}
\tablecaption{Disk and Halo Properties of Edge-on Galaxies\label{tbl-3}}
\tablewidth{0pt}
\tablehead
{
{Galaxy}   & {{\sc H\,i} mass} & $L_{TIR}$/${D_{25}}^2$  & {1C, H, \tablenotemark{a}}  & {Thickest Component} & {{\sc H\,i} Lag} & {Ref.}\tablenotemark{*}\\
{} & \textit{$10^{9}$ $M_{\odot}$}  & {$10^{40}$ $erg s^{-1} kpc^{2}$}  & {or F} & Scale Height (\textit{kpc})  & \textit{km s$^{-1}$ kpc$^{-1}$} & {}
}
\startdata
\phd UGC 1281&2.75 \tablenotemark{b}&n/a\tablenotemark{c}&F&0.18$\rightarrow$0.31&n/a&1, 2\\ 
\phd UGC 7321&1.1&0.35&H, F&2.2 \tablenotemark{d}&$\le|-25|$&3 - 7 \\
\phd NGC 4244&2.5&0.53&1C&0.56&$-9$&3, 5, 8, 9\\
\phd NGC 5746&10&1.16&1C&0.4&n/a&3, 4, 10, 11 \\
\phd NGC 4565&9.9&1.45&F&0.2 $\rightarrow$0.6&$-30$, $-40$&3, 9, 11, 12 \\ 
\phd Milky Way&8&3.0&H, F&1.6&$-22$,$-15$&  11, 13, 14\\
\phd NGC 3044&3.0&6.62&H&0.42 \tablenotemark{d}&n/a&3, 4, 14, 15\\ 
\phd NGC 891&4.1&8.71&H, F&1.25$\rightarrow$2.5&$-15$&3, 8, 12, 16\\ 
\phd NGC 5775&9.1&25.6&1C + features&1&n/a&3, 4, 14, 17\\
\enddata
\tablenotetext{a}{1-component (1C), halo (H), flare (F), or a combination of these.}
\tablenotetext{b}{(Private communication) P. Kamphuis.}
\tablenotetext{c}{Appropriate $L_{TIR}$ is not available for this galaxy.  However, based on H$\alpha$, its star formation rate is 0.0084 M$_{\odot}$ yr$^{-1}$ \citep{2001AJ....121.2003V}, rendering it among the lowest of the galaxies presented here based on calculated rates for the rest. }
\tablenotetext{d}{Gaussian as opposed to exponential layer. In the case of NGC 891, see Oosterloo et al. 2007 for details on the distribution used to fit the halo component.}
\tablenotetext{*}{References: (1) Kamphuis et al. 2011, (2) van Zee 2001 , (3) This work, (4) de Vaucouleurs et al. 1991, (5) Dale et al. 2009, (6) Uson \& Matthews 2003, (7) Matthews \& Wood 2003, (8) Heald et al. 2012, (9) Zschaechner et al. 2011, (10) Moshir et al. 1990, (11) Rand \& Benjamin 2008, (12) Rice et al. 1988, (13) Marasco \& Fraternali 2011, (14) Sanders et al. 2003, (15) Lee \& Irwin 1997, (16) Oosterloo et al. 2007, (17) Irwin 1994}
\end{deluxetable*}

\par  
    The total infrared luminosity per unit area for each galaxy is calculated in this work unless otherwise stated.  These calculations are done based on equations found in \citet{2009ApJ...703..517D}, and involve the optical area (\citealt{1991trcb.book.....D} in most cases; additional references are provided in the table).  Spitzer MIPS data are used if available for a given galaxy, otherwise IRAS data are used. 

\par
    Immediately apparent is the pattern of higher infrared luminosity per unit area (i.e. star formation rate per unit area) and the presence of extra-planar {\sc H\,i} in the form of a halo.  The obvious exception is UGC 7321.  NGC 5775 could be considered an exception, since a global {\sc H\,i} halo is absent.  However, in that case, prominent extra-planar features including extensions and arcs are present. 

\par
    Still, we must emphasize that this result is based on only a few galaxies and modeling methods differ amongst them.  Further analysis of the HALOGAS sample, in which individual galaxies are modeled in a fashion consistent from galaxy to galaxy, will provide a stronger test of this trend.

\subsection{The Lag in NGC 4565 and Other Galaxies}\label{Section6.4}

\par
     We do not detect substantial extra-planar {\sc H\,i} in the form of a global halo, which indicates that there is no strong disk-halo cycling in NGC 4565.  However, the {\sc H\,i} has some vertical velocity dispersion and experiences a lag when it rises above the plane of the disk.  The lag we detect is within the innermost 4.75' (14.9 kpc) and 4.25' (13.4 kpc) in the approaching and receding halves, respectively.  As is the case for UGC 1281 \citep{2011MNRAS.414.3444K} and NGC 4244 \citep{2011ApJ...740...35Z}, this indicates that substantial extra-planar gas need not be present in order for a lag to be observed.  (Although for UGC 1281, a larger line of sight warp component is favored over a lag, yet a lag cannot be ruled out.) 

\par
    The {\sc H\,i} lag detected in NGC 4565 is very steep.  This would be consistent with the \citet{2007ApJ...663..933H} trend in which DIG lags are seen to steepen with decreasing SFR, provided that {\sc H\,i} and DIG lags share a common origin.  Although the DIG lag in NGC 4565 is unknown, modeling of the DIG within the HALOGAS project is currently underway (Wu et al. 2013, \textit{in prep}).  

\par
    A radially varying lag is seen in NGC 4565, which we will now discuss. The onset of this shallowing is at 2.75' (8.6 kpc) in the approaching half and 2.25' (7 kpc) in the receding.  Given the large uncertainty associated with the lag in general, it is difficult to constrain the manner in which it shallows, although outside of 4.75' (14.9 kpc) in the approaching half, and 4.25' (13.4 kpc) in the receding half, it appears to be zero.  

\par
    An entirely geometric interpretation of the shallowing of a lag at large radii, without any consideration of effects such as extra-planar pressure gradients, drag, or magnetic fields, considers the relationship between the rotational velocity and the gravitational potential:  gas rotating high above the plane of the disk is further from the gravitational center than the gas at the same radius in the midplane, resulting in a slower rotational velocity.  This effect is more pronounced at smaller radii, where a certain height above the disk is comparable to the radial distance from the center, resulting in a larger relative change in the gravitational potential, and thus a steeper lag. This effect may be seen in Figure 3 of \citet{2002ApJ...578...98C} and may in part be why a lag is only seen in the inner third of the disk in NGC 4565. However, it should again be noted that such purely ballistic models have under-predicted observed lag magnitudes (\citealt{2006MNRAS.366..449F}, \citealt{2007ApJ...663..933H}), indicating that additional factors must be considered. 

\par
    This radial variation could also be affected by an extra-planar pressure gradient if the cycling gas does not orbit purely ballistically and is subject to hydrodynamic influences. According to \citet{2002ASPC..276..201B}, a pressure gradient directed radially \textit{inward} may cause a \textit{steeper} global lag.   An \textit{outward} pressure gradient may result in a \textit{shallower} global lag, or even an increase in rotation velocity.  If such pressure gradients vary radially themselves, then a radial change in the lag may occur.  Given how little is currently known of halo pressure gradients, as well as the potentially substantial impacts even small pressure gradients may have, it is difficult to say to what degree these will alter the characteristics of any given lag.  

\par
     In {\sc H\,i}, radially shallowing lags are also seen in  NGC 4244 \citep{2011ApJ...740...35Z} and NGC 891 \citep{2007AJ....134.1019O}, and the Milky Way \citet{2011A&A...525A.134M}.  In NGC 4244 the lag decreases from $-$9 km s$^{-1}$ kpc$^{-1}$ to $-$5 km s$^{-1}$ kpc$^{-1}$ and $-$4 km s$^{-1}$ kpc$^{-1}$ in the approaching and receding halves respectively.  This would correspond to a shallowing of 2-2.5 km s$^{-1}$ kpc$^{-2}$ (although, as stated in that paper, and is the case for NGC 4565, the uncertainties in the radial variation are large).  In NGC 891, the gradient in the inner regions is $-$43 km s$^{-1}$ kpc$^{-1}$ and decreases to $-$14 km s$^{-1}$ kpc$^{-1}$ in outer radii, corresponding to a shallowing of 2.5 km s$^{-1}$ kpc$^{-2}$.  In the Milky way, the lag starts with a magnitude of approximately  $-$45 km s$^{-1}$ kpc$^{-1}$ and reaches a value of $-$15 km s$^{-1}$ kpc$^{-1}$ near 4.5 kpc, resulting in a more rapid shallowing of 6.7 km s$^{-1}$ kpc$^{-2}$. The lag in NGC 4565 also decreases rapidly with radius.  Considering the onset of the shallowing of the lag in each half and their respective lag magnitudes (Figure~\ref{fig13}), this corresponds to a shallowing of 6.4 km s$^{-1}$ kpc$^{-2}$ and 4.8 km s$^{-1}$ kpc$^{-2}$ in each half for our optimal models.  Judging by the minimum and maximum lag models, the shallowing could be between 6.4 and 38 km s$^{-1}$ kpc$^{-2}$ in the approaching half (although the latter value is very unlikely as it would imply that a very rapid drop of $-$60 km s$^{-1}$ kpc$^{-1}$ occurs in a very narrow annulus of 1.6 kpc, which does not seem physically realistic) and between 4 and 7.6 km s$^{-1}$ kpc$^{-2}$ in the receding. Thus, the lag in NGC 4565 appears to shallow more rapidly than in NGC 4244 and NGC 891, but at a similar rate to that of the Milky Way. 

\par
   In NGC 4244, the onset of the shallowing of the lag is seen just within the optical radius, whereas the shallowing of the lag in NGC 891 is closer to its center.  In the Milky Way, the lag begins to shallow immediately at the center, and reaches a constant value near 4.5 kpc.  In NGC 4565, the shallowing of the lag begins at approximately 2.5' (7.9 kpc), which is less than half the optical radius of 8.1' (25.5 kpc).  The lag then reaches a constant value of zero near 5' (15.7 kpc).  Thus, there does not yet appear to be a characteristic radius at which the lags begin to shallow, or where the lags reach constant values.

\par
    In addition to these edge-on examples, NGC 4559, a moderately inclined galaxy also shows evidence for a radially decreasing lag \citep{2005A&A...439..947B}.  Although a true velocity gradient in height above the plane of the disk cannot be measured in this case, \citet{2005A&A...439..947B} find a decrease in the rotational velocity of the thick halo component compared to the thin disk of roughly $-$60 km s$^{-1}$ near the center of the galaxy, and $-$20 km s$^{-1}$ at large radii.  As long as the vertical extent of the lagging halo does not fall in a similar way, this drop would indicate a true radially decreasing lag. However, because of its inclination, NGC 4559 does not yield an unambiguous case of a radially varying lag.    
 
\par
     This small sample shows evidence for discrepancies in the nature of the radial variations in lags in each galaxy, including differences in the radius of the onset of shallowing and the rate of shallowing.  Although three edge-on external galaxies as well as the Milky Way clearly show a shallowing lag with increasing radius, a larger sample must be studied in order to extract reliable trends.

\subsection{The Bar in NGC 4565} \label{Section6.5}

\par
\begin{figure}
\epsscale{1}
\figurenum{16}
\plotone{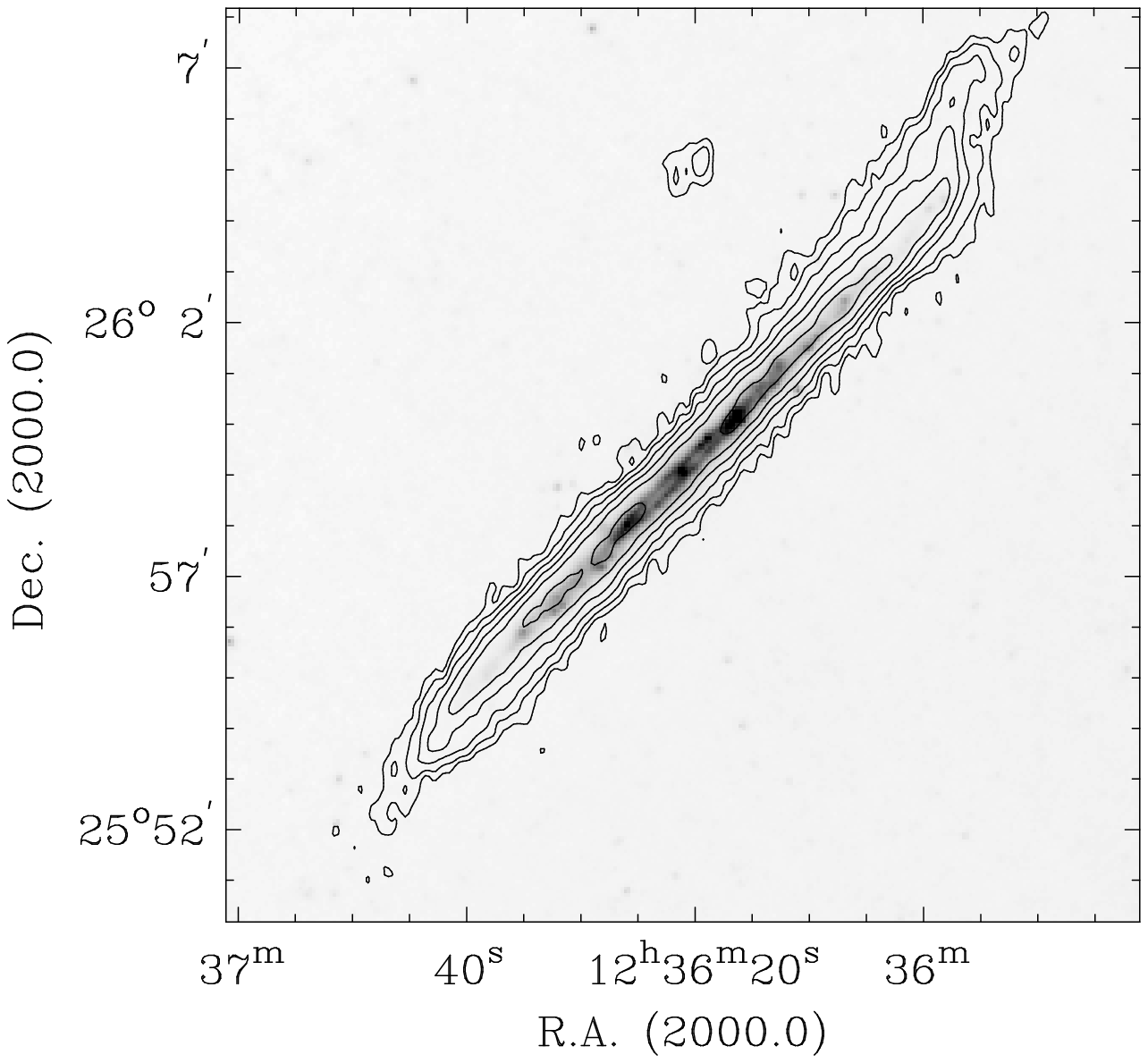}
\caption{Grayscale of 24 $\mu$m MIPS data with {\sc H\,i} contours overlaid starting at $3.0\times10^{19}\mathrm{cm^{-2}}$ and increasing by factors of 2.  Note the prominent ring with a radius of just under 2 arcminutes. \label{fig16}}
\end{figure}

     We find kinematic evidence for radial motions, which we associate with a bar.  Given the inclination of NGC 4565, we only observe a rather limited 2-D projection; not nearly enough to determine its length and thickness using only spatial information, so we must rely on its kinematics.  However, observations at other wavelengths are considered in order to help constrain morphological properties.  In the optical, the bar is seen as a boxy bulge \citep{2010ApJ...715L.176K}.  Additionally, a ring may be seen in higher resolution 24$\mu$ MIPS (NASA/IPAC Infrared Science Archive) data (Figure~\ref{fig16}), as well as CO and 1.2 mm continuum emission (\citealt{1996A&A...310..725N}, Yim et al. 2012 \textit{in prep}).  The radius of the ring (1' to 1.7' in the 24$\mu$ data, and peaking from 1' to 1.5' with an outer radius of ~ 3' in the CO data) is consistent with the ring of higher density seen in {\sc H\,i}, and is thus consistent with the length of the bar we have modeled. 

\par
    Parameters related to the bar are difficult to constrain, thus we will only state what may be reliably extracted from these data.  We can say that the bar exists with a half length of approximately 1.25' (4 kpc).  According to \citet{1996A&A...310..725N} the bar's orientation is between 0$\,^{\circ}$ and $\pm$45$\,^{\circ}$ from end-on. Due to limited resolution and projection effects, we cannot improve upon this.  In fact, by adjusting the rotation curve along the bar and radial motions within the disk, these data allow for an even larger azimuthal range.  

\par
   There appear to be radial motions up to $\pm$60 km s$^{-1}$ within the bar.  For these radial motions we cannot discern between inflow and outflow, largely because we cannot determine the sign of the orientation due to projection effects, although within the length of the bar, a net inflow is expected (e.g.\ \citealt{1992MNRAS.259..328A}).  Furthermore, even without changing the sign of the orientation, changing the direction of radial motions produces too little of an effect at most orientations to determine which is correct.  This may be further complicated by radial motions that vary in magnitude along the bar, of which there is some indication in that further from the center there appears to be less need for such motions, but this is complicated by projection effects and is not enough to make meaningful constraints.  These properties are extremely interconnected, and our spatial resolution limited, hence the large ranges presented here. Fortunately, as may be seen in Figure~\ref{fig10}, the effects from radial motions outside the bar far outweigh any contributions from the bar itself.  Thus, we must stress that the conclusions we draw from our final models that are not directly about the bar are unaffected by the bar itself.

\subsection{Radial Motions in the Disk}  \label{Section6.6}

\par
     Observations of purely axisymmetric inflows, not associated with motions along a bar or spiral structure, have been ambiguous (\citealt{2004ApJ...605..183W} and references therein).  Our work does not break this  ambiguity, but rather provides additional constraints. 

\par
    We see clear indications of a net inflow beginning around the ring of higher density (1.75' or 5.5 kpc), and extending out to 2.75' (8.6 kpc).  The extreme value of this inflow is $-$50$\pm$5 km s$^{-1}$.  Unlike the issues related to constraining the bar, the signatures for this inflow are clear and appear to be unique. There are additional regions in which the models are improved by inflow, but these are almost certainly due to spiral arms.  Since this inflow is associated with the ring, it could be associated with the bar potential. Finally, there exists evidence for a net outflow of 10 km s$^{-1}$ beginning near a radius of 6.75' (21.2 kpc) and extending outward in both halves.  

\par
    The observed effects of the outflow are primarily seen above the plane of the disk at a height of 40 to 80" (2-4 kpc).  A range of inclinations corresponding to a warp component along the line of sight were tested, but none could replicate the asymmetric slant in the data.  Furthermore, if the apparent offset from the plane of the disk were due to projection effects alone, then these would correspond to radii between 50 and 70 kpc, placing them well outside the disk. Therefore, these motions likely do exist above the plane of the disk, rather than being due to such effects.  It is also possible that any outflow could be caused by material being pulled outward due to tidal interactions.  

\par
    Both NGC 4559 \citep{2005A&A...439..947B} and NGC 2403 (\citealt{2000A&A...356L..49S}, \citealt{2002AJ....123.3124F}) have shown similar indications of radial inflows, estimated between 10 and 20 km s$^{-1}$.  In each case, the authors note that the inflow is high above the plane of the disk.  As mentioned above, our data and models indicate that most of the outflow in NGC 4565 is in the thicker, flaring layer.  The inflow however, is close to the plane of the disk and is significantly higher than these values.

\section{Summary and Conclusions} \label{Section7}

\par
    Through our models we determine the following: 

\par
    1. In NGC 4565 we see a flaring layer, but no massive, extended extra-planar {\sc H\,i} layer throughout the galaxy.  There are two companion galaxies, one of which is clearly interacting with NGC 4565 via a bridge of extra-planar material.

\par 
    2.  We detect a lag peaking in magnitude at -40 $^{+5}_{-20}$ km s$^{-1}$ between 1.25 and 4.75' (3.9 and 14.9 kpc) on the approaching half, and -30 $^{+5}_{-30}$  km s$^{-1}$ between 1.25 and 4.25' (3.9 and 13.4 kpc) on the receding half.  Both lags quickly decrease with radius and are zero at radii larger than the upper end of the specified range.

\par
    3.  A bar was previously known to exist within NGC 4565 for which we model a half-length of 1.25' (4 kpc).  Our models indicate radial motions along this bar of 20 to 45 km s$^{-1}$.  A ring of relatively high density is seen near the end of the bar.

\par
    4.  Our models indicate a radial inflow beyond the bar that may be associated with the ring of higher density.  This inflow starts at a radius of 1.75' (5.5 kpc) with a value of $-$50$\pm$5 km s$^{-1}$ and decreases with radius before becoming undetectable outside of 2.75' (8.6 kpc).  A net outflow of 10 km s$^{-1}$ may be present at large radii, most prominently seen 2-3 kpc above the plane of the disk.  The origin of this outflow is unclear.

\par
   At this time, we are still limited to a small number of galaxies for which deep observations exist and detailed models have been made.  Even fewer of these, such as NGC 4565, are edge-on.  However, in edge-ons, a trend is now seen suggesting some connection between the presence of substantial extra-planar {\sc H\,i} and star formation rate per unit area within a given galaxy.  Additionally, we now see radially shallowing lags in three edge-on galaxies, and possibly one that is moderately inclined.  Further work with the full HALOGAS sample will either dismiss or strengthen these early results.

\section{Acknowledgments} \label{Section8}

\par 
     We thank the operators at WSRT for overseeing the observations.  The Westerbork Synthesis Radio Telescope is operated by the ASTRON (Netherlands Institute for Radio Astronomy) with support from the Netherlands Foundation for Scientific Research (NWO).  This research has made use of the NASA/IPAC Infrared Science Archive, which is operated by the Jet Propulsion Laboratory, California Institute of Technology, under contract with the National Aeronautics and Space Administration. We acknowledge the HALOGAS team, especially Paolo Serra for providing a script to create optimal zeroth-moment maps, as well as Erwin de Blok, Filippo Fraternali, Eva J\"{u}tte, Peter Kamphuis and Cat Wu.  Finally, we thank the anonymous referee for his/her insightful comments leading to the improvement of this paper. This material is based on work partially supported by the National Science Foundation under grant AST-0908106 to R.J.R. 

\bibliographystyle{apj}
\bibliography{NGC4565draft}

\nocite{2003ApJ...593..721M}
\nocite{2003AJ....125.2455U}
\nocite{1990IRASF.C......0M}
\nocite{1988ApJS...68...91R}
\nocite{2003AJ....126.1607S}
\nocite{1997ApJ...490..247L}

\end{document}